\documentclass[runningheads]{llncs}
\usepackage{graphicx}
\usepackage{graphicx}
\usepackage{caption}
\usepackage{subcaption}
\usepackage{tikz}
\usetikzlibrary{arrows,backgrounds,calc,decorations.pathmorphing,decorations.pathreplacing,fit,matrix,patterns,petri, positioning, shapes,shapes.multipart,matrix,positioning,shapes.callouts,shapes.arrows,calc, decorations.shapes}
\tikzset{
	silent/.style={
		minimum width{=}2mm,
		minimum height{=}5mm,
		fill{=}gray
	}
}

\usepackage{hyperref}
\usepackage{amssymb}
\usepackage{amsmath}
\usepackage{xcolor,colortbl} 
\usepackage[inline]{enumitem}

\definecolor{logMove}{HTML}{000000}
\definecolor{modelMove}{HTML}{666666}
\definecolor{synchronousMove}{HTML}{d9d9d9}

\usepackage[ruled,lined,vlined]{algorithm2e}

\SetCommentSty{xCommentSty}

\newcommand{\activity}{\ensuremath{a}}
\newcommand{\alignment}{\ensuremath{\gamma}}
\newcommand{\caseId}{\ensuremath{c}}

\newcommand{\event}{\ensuremath{e}}

\newcommand{\labelFunc}{\ensuremath{\lambda}}
\newcommand{\marking}{\ensuremath{M}}
\newcommand{\mset}{\ensuremath{B}}

\newcommand{\petriNet}{\ensuremath{N}}
\newcommand{\place}{\ensuremath{p}}
\newcommand{\places}{\ensuremath{P}}
\newcommand{\pnArcs}{\ensuremath{F}}
\newcommand{\prefixAlignment}{\ensuremath{\overline{\alignment}}}

\newcommand{\proj}{\ensuremath{\pi}}

\newcommand{\sequence}{\ensuremath{\sigma}}
\newcommand{\silent}{\ensuremath{\tau}}

\newcommand{\stream}{\ensuremath{S}}
\newcommand{\transition}{\ensuremath{t}}
\newcommand{\transitions}{\ensuremath{T}}
\newcommand{\univAct}{\ensuremath{\mathcal{A}}}
\newcommand{\univAlign}{\ensuremath{\Gamma}}
\newcommand{\univCase}{\ensuremath{\mathcal{C}}}

\newcommand{\univMSet}{\ensuremath{{\mathcal{B}}}}
\newcommand{\univPrefAlign}{\ensuremath{\overline{\Gamma}}}

\begin{document}
\title{Scalable Online Conformance Checking Using Incremental Prefix-Alignment Computation}
\titlerunning{Scalable Online Conformance Checking}
\author{Daniel Schuster\inst{1}\orcidID{0000-0002-6512-9580} \and
Gero J. Kolhof\inst{2} %\and
%Sebastiaan J. van Zelst\inst{3}
}
\authorrunning{D. Schuster and G. J. Kolhof}
\institute{
    Fraunhofer Institute for Applied Information Technology, Sankt Augustin, Germany\\
    \email{daniel.schuster@fit.fraunhofer.de}\\
    %\url{https://www.fit.fraunhofer.de/} 
    \and
    RWTH Aachen University, Aachen, Germany\\
    \email{gero.kolhof@rwth-aachen.de}
}
\maketitle              % typeset the header of the contribution
\begin{abstract}
Conformance checking techniques aim to collate observed process behavior with normative/modeled process models.
The majority of existing approaches focuses on completed process executions, i.e., offline conformance checking. 
Recently, novel approaches have been designed to monitor ongoing processes, i.e., online conformance checking.
Such techniques detect deviations of an ongoing process execution from a normative process model at the moment they occur.
Thereby, countermeasures can be taken immediately to prevent a process deviation from causing further, undesired consequences.
Most online approaches only allow to detect approximations of deviations.
This causes the problem of falsely detected deviations, i.e., detected deviations that are actually no deviations.
We have, therefore, recently introduced a novel approach to compute exact conformance checking results in an online environment.
In this paper, we focus on the practical application and present a scalable, distributed implementation of the proposed online conformance checking approach.
Moreover, we present two extensions to said approach to reduce its computational effort and its practical applicability.
We evaluate our implementation using data sets capturing the execution of real processes.

\end{abstract}
%
% MAX PAGE LENGTH 15 !!!!!!!!!!!!!!!!!!!!!!!!!!!!!!!!!!!!!!!!!!!!!!!!!!!!!!!!!!!!!!!!!!!!!!!!!!!!!!
%
\section{Introduction}

To achieve operational excellence, accurate knowledge of the different processes executed within one's company is of utmost importance.
Today's information systems accurately track and store the executions of said processes, i.e., \emph{event data}.
The field of \emph{process mining}~\cite{DBLP:books/sp/Aalst16} deals with the analysis of such event data to increase the overall knowledge and insights about the execution of a process.
%Process mining comprises three main areas, i.e., \emph{process discovery}, \emph{conformance checking} and \emph{process enhancement}.
%In process discovery~\cite{DBLP:journals/corr/AugustoCDRMMMS17}, the main aim is to, on the basis of the captured event data, discover a process model that accurately describes the process, i.e., in a (semi-)automated fashion.
%In conformance checking~\cite{DBLP:books/sp/CarmonaDSW18}, the main focus of this paper,  the goal is to assess to what degree a given process model (possibly discovered) effectively describes the captured event data, i.e., allowing us to assess compliance and adherence to predefined policies.
%Finally, process enhancement is concerned with enhancing process models with facts derived from the data, e.g., by calculating and visualizing bottlenecks in the process and decision points.

\begin{figure}[tb]
    \centering
    \includegraphics[width=.91\textwidth,trim={0 9.2cm 4cm 0},clip]{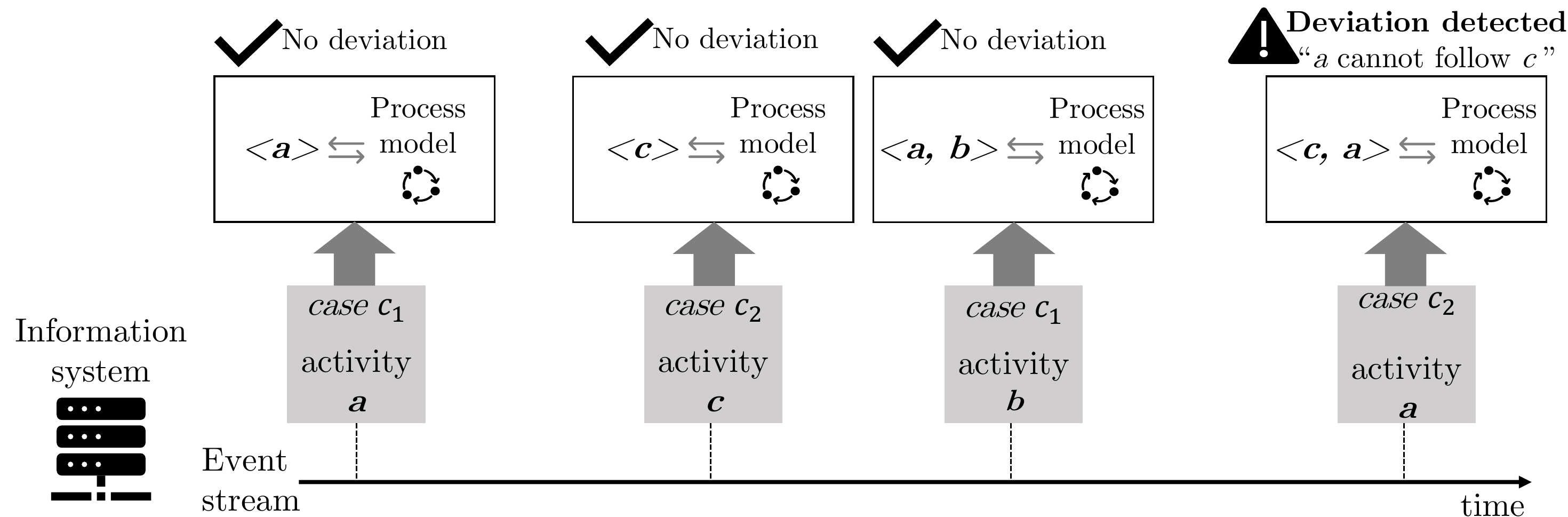}
    \caption{Idea of online conformance checking. Events are observed over time, e.g., activity ``$a$'' was executed for process instance ``$c_1$''. Upon receiving an event, we check if the newly observed event causes a deviation w.r.t. a reference model %Thus, we are able to deduce violations such as ``activity $a$ followed activity $c$ for process $c_2$, which is not allowed''. }
    }
    \label{fig:overview}
\end{figure}

In conformance checking~\cite{DBLP:books/sp/CarmonaDSW18}, the goal is to assess to what degree a given process model describes the captured event data, i.e., allowing us to assess compliance and adherence to predefined policies.
Most approaches in conformance checking focus on offline techniques.
Thus, historical event data is extracted and analyzed.
Whereas this type of analysis is helpful to gain a better understanding of a process' execution and to audit the past execution, it does not allow one to actively intervene in running process executions.
%Interestingly, only a limited amount of work has been devoted to the domain of real-time/online process mining, i.e., analysis of the execution of the process at the moment it unfolds~\cite{DBLP:conf/cec/BurattinSA14,DBLP:conf/bpm/BurattinZADC18,DBLP:journals/kais/ZelstDA18,DBLP:journals/ijdsa/ZelstBHDA19}. %%%% @ DANIEL: IF POSSIBLE EXTEND THIS 
Consider \autoref{fig:overview}, where we visualize the idea of process monitoring.
By applying online conformance checking, we are actively able to detect and pinpoint faulty process executions and communicate such deviations to the process owner.

Work covering the notion of online conformance checking is roughly subdivided into two categories.
First, techniques that indicate (non-)conformance on the basis of abstractions of the process models and/or the event data~\cite{DBLP:conf/bpi/BurattinC17,DBLP:conf/bpm/BurattinZADC18,DBLP:conf/cec/BurattinSA14}.
Secondly, techniques that do not work on abstractions of the input~\cite{DBLP:journals/ijdsa/ZelstBHDA19,schuster_process_monitoring}.
Clearly, abstractions help as a first indicator of (non-)compliance, however, they remain inaccurate, e.g., ``wrongly detected'' deviations.
Whereas the techniques computing exact results enable us to detect and understand non-conformance issues more accurately.
However, for practical applications one needs to resort to approximation schemes because of the computational time involved.
%The downside of such approximation schemes, however, is the potential over-estimation of non-conformity, i.e., ``wrongly detected'' deviations.

In previous work~\cite{schuster_process_monitoring}, we presented an approach for incrementally computing prefix-alignments on event streams, i.e., an exact technique to detect deviations.
Moreover, we showed that the approach outperforms an existing state-of-the-art approximation technique~\cite{DBLP:journals/ijdsa/ZelstBHDA19}.
%Thereby, we assume an event stream and a reference process model as input and calculate optimal prefix-alignment upon receiving a new event.
The main focus of the previous work, however, is on the theoretical foundations.
We showed that we can efficiently compute prefix-alignments on an event stream by continuing a shortest path problem based on an extended search space upon receiving a new event.
In this paper, we focus on the practical application of the presented approach.
Therefore, we present a scalable and distributed implementation of the approach.
In addition, we present extensions that improve the calculation time of the approach to enhance its practical applicability.
Finally, we demonstrate the practical applicability of our approach by applying it to event streams from real-life processes. 

The remainder is structured as follows.
In \autoref{sec:related_work}, we present related work.
In \autoref{sec:preliminaries}, we present preliminaries.
In \autoref{sec:incr_search}, we present an implementation of the incremental prefix-alignment computation and extensions improving the practical applicability by reduced computational effort. 
In \autoref{sec:evaluation}, we evaluate our implementation.
\autoref{sec:conclusion} concludes this paper.

\section{Related Work}
\label{sec:related_work}

Process mining comprises a variety of different techniques such as: process discovery, conformance checking, process enhancement and enrichment techniques. 
For an overview, we refer to \cite{DBLP:books/sp/Aalst16}. 
Next, we focus on conformance checking.

Token-based replay~\cite{DBLP:journals/is/RozinatA08} and footprint comparison~\cite{DBLP:books/sp/Aalst16} are one of the first techniques in the area of conformance checking. 
Both techniques have drawbacks described in \cite{DBLP:books/sp/Aalst16}. 
Therefore, alignments have been introduced \cite{adriansyah_2014_phd_aligning,DBLP:journals/widm/AalstAD12} that map traces onto an acceptable path of a given process model. 
Moreover, alignments indicate mismatches between observed and modeled behavior.
The problem of finding an alignment was shown to be reducible to a shortest path problem \cite{adriansyah_2014_phd_aligning,adriansyah2013memory}.  

The previously mentioned techniques are designed for offline usage. 
Thus, deviations can only be detected post-mortem, i.e., after the process instance has already finished.
In \cite{DBLP:journals/ijdsa/ZelstBHDA19} an approach was presented to monitor ongoing process executions based on event streams. 
Essentially, a framework was introduced that computes prefix-alignments for ongoing processes each time a new activity has been performed. 
Moreover, the framework includes multiple options to decrease computational effort; in return, false negatives in terms of deviation detection may arise. 
In \cite{schuster_process_monitoring} it was shown that prefix-alignments can be computed in an online setting by continuing a shortest path search on an extended search space upon receiving a new event. 
Moreover, this approach guarantees optimal prefix-alignments.
Thus, false deviations w.r.t. deviation detection cannot occur.

In \cite{DBLP:conf/bpm/BurattinZADC18} an approach was presented that computes the conformance on event streams, too.
In contrast to the prefix-alignment approaches, conformance of a process execution is computed based on behavioral patterns that describe control flow relations between activities.
Furthermore, this approach is suited for partial and for already running process executions where past information on such process executions is not available. 
In return, it uses abstractions, i.e., behavioral patterns, of the process model and the event log that leads to a loss of expressiveness in the deviation explanation. 
Another approach calculates an extended transition system for a given process model in advance. 
Such extended transition system allows for replaying the ongoing process \cite{DBLP:conf/bpi/BurattinC17}. 
Costs are attached to the edges in an extended transition system, and replaying a divergent, non-compliant process instance leads to costs greater than zero.

\iffalse
In addition to the presented approaches, some general work on process mining on event streams has been done by \cite{DBLP:reference/bdt/Burattin19}. 
Various, general applicable, high-level strategies are presented that deal with issues such as, how to handle the limited amount of storage capabilities and the potential infinite event stream in an online scenario. 
In addition, approaches for efficiently storing event data and filtering event streams have been presented in \cite{771d457cde05489da00b3c863e01596d}. 
\fi

\section{Preliminaries}
\label{sec:preliminaries}

A multiset $\mset$ over a set $X$ contains an arbitrary number of each element in $X$. 
%Formally, a multiplicity is assigned to each element in $X$, i.e., $\mset \colon X  {\to} \naturalsZero$.
%We write a multiset $\mset$ as $[x^i,y^k,z^n]$, where $\mset(x){=}i$, $\mset(y){=}k$ etc.
%In case $\mset(x){=}1$, we omit $x$'s superscript from the multiset notation. 
%If $\mset(x){=}0$, we omit $x$ from the multiset notation.
The set of all possible multisets over a set $X$ is denoted by $\univMSet(X)$.
For instance, $[x^5,y] {\in} \univMSet \big( \{x,y,z\} \big)$ contains $5$ times $x$, once $y$ and no $z$. 
%For a given multiset $\mset$, we write $x {\in_+} B$ if $x$ is contained at least once in $B$, e.g., $x {\in_+} [x^5,y]$.
A sequence $\sequence$ of length $n$ over a set $X$ assigns an element to each index in $\{1,\dots,n\}$, i.e., $\sequence \colon \{1,\dots,n\} {\to} X$.
%The empty sequence is denoted by $\langle\rangle$.
We let $|\sequence|$ denote the length of $\sequence$.
%We write a sequence $\sequence$ as $\langle \sequence(1), \sequence(2), ..., \sequence(|\sequence|)\rangle$.
%Concatenation of two sequences $\sequence$ and $\sequence'$ is written as $\sequence {\cdot} \sequence'$, e.g., $\langle x,y\rangle {\cdot} \langle z \rangle {=} \langle x,y,z \rangle$.
The set of all possible sequences over a set $X$ is written as $X^*$, e.g., $\langle a,a,b \rangle {\in}  \{a,b,c,d\}^*$.
%We overload the notation of element inclusion for sequences, i.e., given $\sequence{\in}X^*$ and $x{\in}X$, we write $x{\in}\sequence$ if $\exists_{1\leq i\leq|\sequence|}\left(\sequence(i)=x\right)$, e.g., $b {\in} \langle a,b\rangle$. 

\iffalse
Next, we introduce projection functions.
Let $\sequence{\in}X^*$ and $X'{\subseteq}{X}$, we define $\sequence_{\downarrow_{X'}}{\in}X'^*$ with:
$\langle\rangle_{\downarrow_{X'}} {=} \langle\rangle$, $(\langle x\rangle\cdot\sequence)_{\downarrow_{X'}} {=}x \cdot\sequence_{\downarrow_{X'}}$ if $x{\in}X'$ and $(\langle x\rangle\cdot\sequence)_{\downarrow_{X'}} {=} \cdot\sequence_{\downarrow_{X'}}$ if $x{\notin}X'$.
Let $t{=}(x_1,x_2,...,x_n) {\in} X_1 {\times} X_2 {\times} \cdots {\times} X_n$ be a $n$-tuple, we let $\proj_1(t){=}x_1,\allowbreak \proj_2(t){=}x_2 , \allowbreak \dots,\allowbreak \proj_n(t){=}x_n$ denote the corresponding projection functions that extract a specific component of the given tuple.
Correspondingly, given a sequence $\sequence$ with length $m$ of $n$-tuples  $\sequence{=}\langle(x^1_1,...,x^1_n),\allowbreak \dots,\allowbreak (x_1^m,...,x_n^m)\rangle$, we define the corresponding projection functions $\proj^*_1(\sequence){=}\langle x_1^1,\dots,x_1^m\rangle,\allowbreak \dots,\allowbreak \proj^*_n(\sequence){=}\langle x_n^1,...,x_n^m\rangle$ that extract a specific component, i.e., a $n$-tuple, of the given sequence.
\fi

\subsection{Event Data and Event Streams}

Today's information systems deployed in organizations capture the execution of (business) processes in great detail.
These systems record the executed activity, the corresponding process instance within the activity was executed and potentially many other attributes. 
We refer to such data as \emph{event data}.

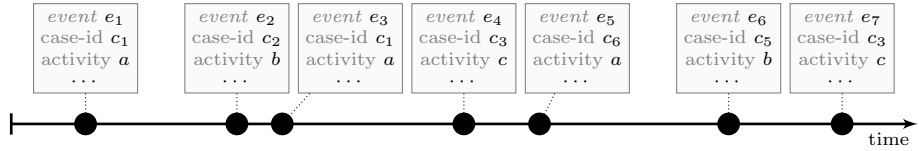
\begin{figure}[tb]
    \centering
    \scriptsize
    \resizebox{\textwidth}{!}{%
    \begin{tikzpicture}
    \tikzstyle{e}=[draw=gray,fill=black!2]
        
        \node[e,align=center,font=\scriptsize] (e1) at (1,1) {\textit{\textcolor{gray}{event} $e_1$}\\\textcolor{gray}{case-id} $c_1$ \\\textcolor{gray}{activity} $a$\\$\cdots$};
        \draw[densely dotted] (e1) to (1,0);
        
        \node[e,align=center,font=\scriptsize] (e2) at (3,1) {\textit{\textcolor{gray}{event} $e_2$}\\\textcolor{gray}{case-id} $c_2$ \\\textcolor{gray}{activity} $b$\\$\cdots$};
        \draw[densely dotted] (e2) to (3,0);
        
        \node[e,align=center,font=\scriptsize] (e3) at (3.6+.9,1) {\textit{\textcolor{gray}{event} $e_3$}\\\textcolor{gray}{case-id} $c_1$ \\\textcolor{gray}{activity} $a$\\$\cdots$};
        \draw[densely dotted] (e3) to (3.6,0);
        
        \node[e,align=center,font=\scriptsize] (e4) at (6,1) {\textit{\textcolor{gray}{event} $e_4$}\\\textcolor{gray}{case-id} $c_3$ \\\textcolor{gray}{activity} $c$\\$\cdots$};
        \draw[densely dotted] (e4) to (6,0);
        
        \node[e,align=center,font=\scriptsize] (e5) at (7+.5,1) {\textit{\textcolor{gray}{event} $e_5$}\\\textcolor{gray}{case-id} $c_6$ \\\textcolor{gray}{activity} $a$\\$\cdots$};
        \draw[densely dotted] (e5) to (7,0);
        
        \node[e,align=center,font=\scriptsize] (e6) at (9.5,1) {\textit{\textcolor{gray}{event} $e_6$}\\\textcolor{gray}{case-id} $c_5$ \\\textcolor{gray}{activity} $b$\\$\cdots$};
        \draw[densely dotted] (e6) to (9.5,0);
        
        \node[e,align=center,font=\scriptsize] (e7) at (11,1) {\textit{\textcolor{gray}{event} $e_7$}\\\textcolor{gray}{case-id} $c_3$ \\\textcolor{gray}{activity} $c$\\$\cdots$};
        \draw[densely dotted] (e7) to (11,0);        
        
        \draw[line width=1pt, |->, >=latex'](0,0) -- coordinate (x axis) (12,0) node[below left] {time}; 
        \fill[black] (1,0) circle (0.15);
        \fill[black] (3,0) circle (0.15);
        \fill[black] (3.6,0) circle (0.15);
        \fill[black] (6,0) circle (0.15);
        \fill[black] (7,0) circle (0.15);
        \fill[black] (9.5,0) circle (0.15);
        \fill[black] (11,0) circle (0.15);
       
    \end{tikzpicture}
    }
    \caption{Visualization of an event stream \iffalse . Dots represent events containing the label of the executed activity, the process instance/case-id, a timestamp and potentially various other attributes\fi}
    \label{fig:example_stream}
\end{figure}

In this paper, we assume an (infinite) event stream. 
Each event contains information as described above, i.e., it describes the execution of an activity within a process instance/case. 
In this paper, however, we are only interested in the label of the executed activity, the case-id of the corresponding process instance and the order of events.
Consider \autoref{fig:example_stream} for an example event stream.
%Note that at any given time, our knowledge of a process instance is incomplete since a new event may occur for it at any later time.
Next, we formally define an event stream.

\begin{definition}[Event; Event Stream]
\label{def:event_stream}
Let $\univCase$ denote the universe of \emph{case identifiers} and $\univAct$ the universe of \emph{activities}. 
An event $\event$ describes the execution of an activity $\activity{\in}\univAct$ in the context of a process instance identified by $\caseId{\in}\univCase$.
An event stream $\stream$ is a sequence of events, i.e., $\stream{\in}(\univCase{\times}\univAct)^*$.
\end{definition}

\subsection{Process Models}

Process models allow us to describe process behavior.
In this paper, we focus on \emph{sound Workflow nets}~\cite{DBLP:journals/jcsc/Aalst98}.
Workflow nets (WF-nets) are a subclass of \emph{Petri nets}~\cite{murata_1989} %that allow us to represent concurrent behavior in a compact manner. 
and sound WF-nets are a subclass of WF-nets with preferred \emph{behavioral properties} guaranteeing the absence of deadlocks, livelocks and other anomalies.

\begin{figure}[b]
	\centering
    	\footnotesize
    	    \resizebox{.4\textwidth}{!}{%

    	\begin{tikzpicture}[node distance=1.4cm,>=stealth',bend angle=20,auto, label distance=-.06cm]
     		\tikzstyle{place}=[circle,thick,draw=black,minimum size=5mm]
    	  	\tikzstyle{transition}=[thick,draw=black,minimum size=5mm]
    	  	\tikzstyle{silent}=[rectangle,thick,draw=black,fill=black,minimum size=5mm, text=white]
    	    \node [place,tokens=1,label=below:{$p_1$}] (p1) {};
    		\node [transition] (A) [right of = p1, label=below:{$t_1$}, label={[align=center,color=gray]above:create account}]  {$a$};    
    	    \node [place] (p2) [right of = A,label=below:{$p_2$}] {};
    		\node [silent] (tau) [below of =A, label=above:{$t_2$}]  {$\tau$};
    		\node [transition] (B) [right of= p2,label=below:{$t_3$},label={[align=center,color=gray]above:submit order}]  {$b$};
    		\node [transition] (C) [below of= B,label=above:{$t_4$},label={[align=center,color=gray]below:request quote}]  {$c$};
    	    \node [place] (p3) [right of= B,label=below:{$p_3$}] {};
    		\draw [->] (p1) to (A); 
    		%\draw [->, bend left] (A) to (p2); 
    		\draw [->] (A) to (p2); 
    		\draw [->] (p2) to (B); 
    		\draw [->] (p2) to (C); 
    		\draw [->] (B) to (p3); 
    		\draw [->] (C) to (p3); 
    		
    		\draw [->] (p1) to (tau);
    		\draw [->] (tau) to (p2);
	    \end{tikzpicture}
	    }
	\caption{Example WF-net $\petriNet_1$ with visualized initial marking $\marking_i{=}[\place_1]$ and final marking $M_f{=}[\place_3]$ describing a simplified ordering process}
	%starting with an optional activity \emph{create account}, followed by a choice of either \emph{submit order} or \emph{request quote}}
	\label{fig:example_petri_net}
\end{figure}
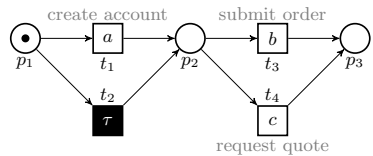

A Petri net $\petriNet {=} (\places,\transitions,\pnArcs, \labelFunc)$ consists of a set of \emph{places} $\places$, \emph{transitions} $\transitions$ and \emph{arcs} $\pnArcs {=} (\places{\times}\transitions){\cup}\allowbreak(\transitions{\times}\places)$ connecting places and transitions.
%However, an arc never directly connects two transitions or two places.
 %, where $\pnArcs {=} (\places{\times}\transitions)\cup(\transitions{\times}\places)$ describes the arcs of the Petri net.
%For example, in \autoref{fig:example_petri_net}, $\pnArcs{=}\{(\place_i,\transition_1),(\transition_1,\place_2),\dots,(\transition_4,\place_3)\}$.
Given the universe of activities $\univAct$, the \emph{labeling function} $\lambda{\colon}\transitions{\to}\allowbreak\univAct{\cup}\{\tau\}$ assigns an (possibly invisible, i.e., $\tau$) activity label to each transition. For instance, $\labelFunc(t_1){=}a$ and $\labelFunc(t_2){=}\silent$ (\autoref{fig:example_petri_net}). 

A state of a Petri net is defined by its \emph{marking} $\marking$ that is defined as a multiset of places, i.e. $\marking{\in}\univMSet(\places)$.
%Graphically we represent a marking by drawing black dots, i.e. in the case of $\marking(\place){=}k$ we draw $k$ dots in place $p$.
%Observe that the visualized marking of the Petri net $N_1$ (\autoref{fig:example_petri_net}) is $[\place_1]$.
Given a Petri net $\petriNet$ and a marking $\marking$, a \emph{marked net} is written as $(\petriNet,\marking)$.
We write $\marking_i$/$\marking_f$ to represent the initial/final marking.

For $x{\in}\places {\cup} \transitions$, we define the set of all elements having an incoming arc from $x$, i.e., $x{\bullet}{=}\left\{y {\in}\places {\cup} \transitions \mid (x,y){\in}\pnArcs\right\}$.
Symmetrically, we define ${\bullet} x {=} \left\{y{\in}\places{\cup}\transitions \mid (y,x){\in}\pnArcs\right\}$.
Transitions allow for changing the state of a Petri net.
Given a marking $\marking{\in}\univMSet(\places)$, we call a transition $t$ \emph{enabled} if all incoming places contain at least one token, i.e., $\forall \place{\in}{\bullet}\transition\left(\marking(\place){>}0\right)$.
We write $(\petriNet,\marking)[\transition\rangle$ if $\transition$ is enabled in marking $\marking$.
An enabled transition can be \emph{fired}. 
Such firing leads to a state change, i.e., a new marking $\marking'{\in}\univMSet(\places)$, where $\marking'(\place){=}\marking(\place){+}1$ if $\place{\in}\transition{\bullet} {\setminus} {\bullet} \transition$, $\marking'(\place){=}\marking(\place){-}1$ if ${\bullet} \transition {\setminus} \transition {\bullet}$ and $\marking'(\place){=}\marking(\place)$ otherwise. 
%Informally, this means that if $t$ is fired, a token is consumed/removed from all incoming places of $t$ and a token is produced/put in each outgoing place of $t$.
%We write $(\petriNet,\marking) {\xrightarrow{\transition}} (\petriNet,\marking')$ to denote that firing the transition $\transition$ in marking $\marking$ leads to marking $\marking'$.
%We overload the notation for sequences of transitions.
Given a sequence of transitions $\sequence{\in}\transitions^*$, we write $(\petriNet,\marking) {\xrightarrow{\sequence}} (\petriNet,\marking')$ to denote that firing the transitions in $\sequence$ leads to $\marking'$. 
%We write $(\petriNet,\marking) {\rightsquigarrow} (\petriNet,\marking')$ if there exists a sequence of transitions $\sequence{\in}T^*$ s.t. $(\petriNet,\marking) {\xrightarrow{\sequence}}(\petriNet,\marking')$.
%We define the state space of a Petri net $N$ with $\marking_i$, i.e., the set of reachable markings, by $\reachableMarkings(\petriNet,\marking_i){=}\{\marking'{\in}\univMSet(\places) \mid (\petriNet,\marking) {\rightsquigarrow} (\petriNet,\marking') \}$.

A \emph{WF-net} is a Petri net with a unique source place $\place_i$, i.e. $\bullet\place_i{=}\emptyset$ and sink place $\place_o$, i.e. $\place_o{\bullet}{=}\emptyset$ that form the initial/final marking, i.e., $\marking_i{=}[\place_i]$/$\marking_f{=}[\place_o]$.
Moreover, all transitions and places are on a path from source to sink.

\subsection{Alignments}

\begin{figure}[tb]
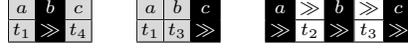

    \newcolumntype{l}{>{\columncolor{logMove}\color{white}}c}
    \newcolumntype{s}{>{\columncolor{synchronousMove}}c}
    \newcolumntype{m}{>{\columncolor{modelMove}\color{white}}c}
    \centering
    \scriptsize
        {
        \begin{tabular}{|s | l | s|}\hline
        $a$     & $b$     & $c$     \\ \hline
        $t_1$   & $\gg$     & $t_4$   \\ \hline
        \end{tabular}
        \hspace{0.4cm}
        \begin{tabular}{|s | s | l|}\hline
        $a$     & $b$     & $c$     \\ \hline
        $t_1$   & $t_3$     & $\gg$   \\ \hline
        \end{tabular}
        \hspace{0.4cm}
        \begin{tabular}{|l | c | l | c | l|}\hline
        $a$     & $\gg$     & $b$   & $\gg$ & $c$   \\ \hline
        $\gg$   & $t_2$     & $\gg$ & $t_3$ & $\gg$  \\ \hline
        \end{tabular}
        }
    \caption{Three possible alignments for $N_1$ (\autoref{fig:example_petri_net}) and the trace $\langle a,b,c \rangle$}
    \label{fig:alignments}
\end{figure}

\emph{Alignments}~\cite{adriansyah_2014_phd_aligning} allow to compare observed behavior with modeled behavior. 
%Thus, the main use case of alignments is the detection of deviations between a trace and a process model.
%Consider the trace $\langle a,b,c\rangle$ and the WF-net $N_1$.
Consider \autoref{fig:alignments} in which we present three possible alignments for the trace $\langle a,b,c\rangle$ and the WF-net $N_1$.
The first row of an alignment (ignoring the skip symbol $\gg$) corresponds to the trace, and the second row corresponds to a sequence of transitions (ignoring $\gg$) leading from the initial to the final marking.
Each column represents an alignment-\emph{move}.
We distinguish three move types.
A \emph{synchronous move} (gray) matches an observed activity to the execution of a transition, e.g., the first move of the first alignment (\autoref{fig:alignments}) with $\lambda(t_1) {=} a$. 
\emph{Log moves} (black) indicate that an observed activity is not re-playable in the current state of the process model. 
%Thus, log moves represent deviations.
\emph{Model moves} (white) indicate that the execution of a transition cannot be mapped onto an observed activity.
We further differentiate \emph{invisible} and \emph{visible model moves}. 
An invisible model move consists of an invisible transition, e.g., for the first model move in the third alignment (\autoref{fig:alignments}) $\labelFunc(\transition_2) {=} \tau$.
Invisible model moves do not represent a deviation.
In contrast, visible model represent a deviation, e.g., the second model move of the third alignment.

Next to alignments, there is the concept of prefix-alignments, which are a relaxed version of conventional alignments. The first row of a prefix-alignment (ignoring $\gg$) also corresponds to the trace, but the second row corresponds to a sequence of transitions (ignoring $\gg$) leading from the initial marking to a marking from which the final marking can be still reached.
Hence, prefix-alignments are suited to compare ongoing processes with a reference model.

\iffalse
\begin{definition}[(Prefix-)Alignment]
Let $\univAct$ be the universe of activities, $\sequence{\in}\univAct^*$, $\petriNet{=}(\places,\transitions,\pnArcs,\place_i,\place_o,\labelFunc)$ a sound WF-net and $\gg {\notin} \univAct{\cup}\transitions$.
A sequence \\$\alignment{\in}\left((\univAct{\cup}\{\gg\})\times(\transitions{\cup}\{\gg\})\right)^*$ is an alignment iff:
\begin{enumerate}
    \item $\sequence{=}\proj^*_1(\alignment)_{\downarrow_{\univAct}}$
    \item  $(N,[\place_i])\xrightarrow{\proj^*_2(\alignment)_{\downarrow_{\transitions}}}(N,[\place_o])$
    \item $(\gg,\gg) {\notin} \alignment$
\end{enumerate}
Similarly, a sequence $\prefixAlignment{\in}\left((\univAct{\cup}\{\gg\})\times(\transitions{\cup}\{\gg\})\right)^*$ is a \emph{prefix-alignment} if the second requirement is relaxed to:
$(N,[\place_i])\xrightarrow{\proj^*_2(\alignment)_{\downarrow_{\transitions}}}(N,\marking) {\rightsquigarrow} (N,[\place_o])$.
\iffalse For a trace $\sequence$ and a WF-net $\petriNet$ let $\univAlign(\sequence,\petriNet)$ denote the set of all possible alignments. Analogously, let $\univPrefAlign(\sequence,\petriNet)$) denote the set of all prefix-alignments.\fi 
\end{definition}
\fi

\begin{figure}[tb]
    \centering
    \begin{subfigure}[t]{0.39\textwidth}
        \centering
        \resizebox{.75\textwidth}{!}{% <------ Don't forget this %
        \begin{tikzpicture}[node distance=1cm,>=stealth',bend angle=45,auto]
        \scriptsize
    	  \tikzstyle{place}=[circle,thick,draw=black,minimum size=4mm]
    	  \tikzstyle{place_new}=[circle,very thick,draw=lightgray,minimum size=4mm,densely dotted]
    
    	  \tikzstyle{transition}=[thick,draw=black,minimum size=4mm]
    	  \tikzstyle{transition_new}=[very thick,draw=lightgray,minimum size=4mm,densely dotted]
    	  \tikzstyle{silent}=[rectangle,very thick,draw=black,fill=black,minimum size=4mm, text=white]
    	
    	    % trace net
    	    \node [place,tokens=1,label={$p_0'$}] (p0_) {};
    		\node [transition] (t1_) [right of = p0_, %label={$(t_1',\gg)$},
    		fill=logMove]  {$\color{white}(a,\gg)$};  
    		\node [place,tokens=0,label={$p_1'$}, right of = t1_] (p1_) {};
    		\node [transition_new] (t2_) [right of = p1_, %label={$(t_2',\gg)$},
    		fill=logMove]  {$\color{white}(b,\gg)$};  
    		\node [place_new,tokens=0,label={$p_2'$}, right of = t2_] (p2_) {};
    		%\node [transition] (t3_) [right of = p2_, label={$(t_3',\gg)$}]  {$(c,\gg)$};  
    		%\node [place,tokens=0,label={$p_3'$}, right of = t3_] (p3_) {};
    		\draw [->,] (p0_) to (t1_); 
    		\draw [->,] (t1_) to (p1_); 
    		\draw [->,lightgray,densely dotted,very thick] (p1_) to (t2_); 
    		\draw [-> ,lightgray,densely dotted,very thick] (t2_) to (p2_); 
    		%\draw [->,] (p2_) to (t3_); 
    		%\draw [->,] (t3_) to (p3_);
    		% synchronous part
    		\node [transition,%label={$(t_1',t_1)$}
    		, below of = t1_,fill=synchronousMove] (t_aa) {$(a,a)$};
    		\draw [->,] (p0_) to (t_aa); 
    		\draw [->,] (t_aa) to (p1_); 
    
    		\node [transition_new,%label={$(t_2',t_3)$}
    		, below of = t2_,fill=synchronousMove] (t_bb) {$(b,b)$};
    		\draw [->,lightgray,densely dotted,very thick] (p1_) to (t_bb); 
    		\draw [->,lightgray,densely dotted,very thick] (t_bb) to (p2_); 
    		
    		%\node [transition,label={$(t_3',t_3)$}, below of = t3_] (t_cc) {$(c,c)$};
    		%\draw [->,] (p2_) to (t_cc); 
    		%\draw [->,] (t_cc) to (p3_); 
    		
    		\node [place,label={$p_1$}, below = 1.7cm of  p0_,tokens=1] (p1) {};
    		\node [transition] (A) [right of = p1, %label={$( \gg,t_1)$}
    		]  {$(\gg,a)$};    
    	    \node [place] (p2) [right of = A,label={$p_2$}] {};
    		\node [transition] (B) [right of=p2,%label={$( \gg,t_3)$}
    		]  {$(\gg,b)$};
    		\node [transition] (C) [below of = B,%label={$(\gg,t_4)$}
    		]  {$(\gg,c)$};
    		\node [transition] (tau) [below of = A,%label={$(\gg,t_2)$}
    		]  {$(\gg,\tau)$};
    
    	    \node [place] (p3) [right of=B,label={$p_3$}] {};
    		\draw [->] (p1) to (A); 
    		%\draw [->, bend left] (A) to (p2); 
    		\draw [->] (A) to (p2); 
    		\draw [->] (p2) to (B); 
    		\draw [->] (p2) to (C); 
    		\draw [->] (B) to (p3); 
    		\draw [->] (C) to (p3); 
    		
    		\draw [->] (p1) to (t_aa);
    		\draw [->] (t_aa) to (p2);
    		
    		\draw [->,lightgray,densely dotted,very thick] (p2) to (t_bb);
    		\draw [->,lightgray,densely dotted,very thick] (t_bb) to (p3);
    		
    		\draw [->] (p1) to (tau); 
    		\draw [->] (tau) to (p2); 
    		
    		%\draw [->] (p2) to (t_cc);
    		%\draw [->] (t_cc) to (p3);
	    \end{tikzpicture}
	    }
        \caption{SPN}
        \label{fig:sync_net_ext_spn}
    \end{subfigure}
    \hfill
    \begin{subfigure}[t]{0.59\textwidth}
        \centering
        \resizebox{.75\textwidth}{!}{% <------ Don't forget this %
        \begin{tikzpicture}[node distance=1.8cm,>=stealth',bend angle=19,auto]
    	\scriptsize
        \tikzstyle{gray}=[rectangle,draw=black,thick,minimum size=4mm]
      
        \tikzstyle{every label}=[blue]
    
        % First net
        \node [gray,tokens=0] (1) {$[p_0',p_1]$};
        \node [gray,tokens=0] (2)[right of = 1] {$[p_1',p_2]$};
        \node [gray,tokens=0] (3)[above of= 2] {$[p_1',p_1]$};
    		
        \node [gray,tokens=0, color=gray,densely dotted] (31)[right of= 3] {$[p_2',p_1]$};
    
        \draw [->,color=gray,densely dotted,thick] (3) to node  {$(b,\gg)$} (31);
        \draw [->] (3) to node [right = -0.1cm,text=black] {$(\gg,a)$} (2);
        
        \node [gray,tokens=0] (4)[below of= 2] {$[p_0',p_2]$};
        
        \node [gray,tokens=0] (444)[right of= 4] {$[p_0',p_3]$};
    	\draw [->,bend left] (4) to node  {$(\gg,b)$} (444); 
    	\draw [->,bend right] (4) to node  {$(\gg,c)$} (444); 
    
        \node [gray,tokens=0] (4444)[above right=1cm of 444] {$[p_1',p_3]$};
    	\draw [->] (444) to node [right] {$(a,\gg)$} (4444);

        \node [gray,tokens=0,color=gray,densely dotted] (5)[right of =2] {$[p_2',p_3]$};
       	\draw [->,color=gray,densely dotted,thick] (4444) to node [right] {$(b,\gg)$} (5); 
    
        \node [gray,tokens=0,color=gray,densely dotted] (6)[right of = 31] {$[p_2',p_2]$};
    	
    	\draw [->] (1) to node [text=black] {$(a,a)$} (2); 
    	\draw [->] (1) to node [left,text=black] {$(a,\gg)$} (3);
    	\draw [->] (1) to node [right,text=black] {$(\gg,a)$} (4);  
    	\draw [->, bend right= 45] (1) to node [left,text=black] {$(\gg,\tau)$} (4);

    	\draw [->,color=gray,densely dotted,thick] (2) to node [] {$(b,b)$} (5);  
    	\draw [->,color=gray,densely dotted,thick] (2) to node [left] {$(b,\gg)$} (6);  
    
        \draw [->, bend left=45,color=gray,densely dotted,thick] (6) to node [right] {$(\gg,b)$} (5);
        \draw [->,color=gray,densely dotted,thick] (6) to node [right] {$(\gg,c)$} (5);
        \draw [->,color=gray,densely dotted,thick] (31) to node {$(\gg,a)$} (6); 
        \draw [->] (4) to node [right=-.1cm,text=black] {$(a,\gg)$} (2); 
        
     	\draw [->] (2) to node [below] {$(\gg,b)$} (4444); 
    	\draw [->,bend right] (2) to node [below] {$(\gg,c)$} (4444);

        \end{tikzpicture}
        }
        \caption{SPN's state space/search-space for shortest path search}
        \label{fig:state_space_spn}
    \end{subfigure}
    \caption{SPN of $\langle a, b \rangle$ and WF-net $\petriNet_1$ (\autoref{fig:example_petri_net})\iffalse and its corresponding state space\fi}
    \label{fig:sync_net_ext}
\end{figure}
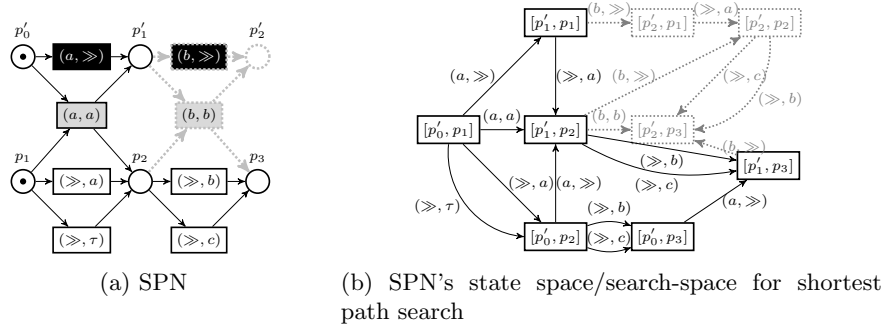

In general, we aim to minimize log and visible model moves.
Since multiple (prefix-)alignments exist for a given model and trace, costs are assigned to alignment moves.
The \emph{standard cost function} assigns cost 0 to synchronous and invisible model moves, and cost 1 to log and visible model moves.
A \mbox{(prefix-)}alignment with minimal costs is called \emph{optimal}.
The computation of an optimal \mbox{(prefix-)} alignment is reducible to a shortest path problem~\cite{adriansyah_2014_phd_aligning}.
Therefore, a \emph{synchronous product net} (SPN) is created for a given trace and WF-net, e.g., as in \autoref{fig:sync_net_ext}.
For a formal definition of a SPN, we refer to~\cite{adriansyah_2014_phd_aligning}.
Note that each transition corresponds to a (prefix-)alignment move.
\autoref{fig:state_space_spn} shows the corresponding SPN's state space, on which the shortest path search is executed. 
Note that each edge represents an alignment move and, hence, has assigned costs.
In the given example, the initial state/marking for the shortest path search is $[p_0',p_1]$ and the goal states are all states/markings containing $p_2'$.

\section{Incremental Prefix-Alignment Computation}
\label{sec:incr_search}
In this section, we initially present the main idea of incrementally computing prefix-alignments on an event stream.
Next, we introduce the proposed implementation of our approach.
Subsequently, we present two extensions to improve the practical applicability of the approach.

\subsection{Background}
\label{sec:incr_search_basics}
The core idea of incrementally computing prefix-alignments is to continue a shortest path search on an extended search space upon receiving a new event.
For each process instance, we cache its current SPN and extend it upon receiving a new event.
Moreover, we store intermediate search-results and reuse them when we continue the search.
Next, we briefly list the main steps of the algorithm.
Assume an event stream and a reference process model as input.

\begin{enumerate}
    \item We receive a new event. For instance, consider $(c_1, b)$ describing that the activity $b$ was executed for the process instance identified by case-id $c_1$.
    
    \item We extend the SPN for process instance $c_1$ by the new activity $b$. For example, assume that we previously received the event $(c_1,a)$. Consider \autoref{fig:sync_net_ext_spn} showing the extension of the SPN highlighted by dashed gray elements. Note that when extending a SPN by a new activity, new transitions are added representing a log move on the new activity and potential synchronous moves.

    \item We continue the search for a shortest path on the state space of the extended SPN from previously cached intermediate search-results, i.e., states already explored/investigated. %Technically, we incrementally execute the A* search algorithm on an extended search space with pre-filled open- and closed set obtained from the previous search. 

    \item We return the prefix-alignment and cache the search-results.
\end{enumerate}

For a detailed overview, we refer to~\cite{schuster_process_monitoring}.
Next, we introduce the proposed implementation of the incremental prefix-alignment computation approach.
%Next, we introduce an extension to avoid solving trivial incremental shortest path problems.

\subsection{Implementation}
\label{sec:implementation}

\begin{figure}[tb]
    \centering
    \includegraphics[width=1\textwidth,clip,trim=0cm 0cm 0cm .1cm]{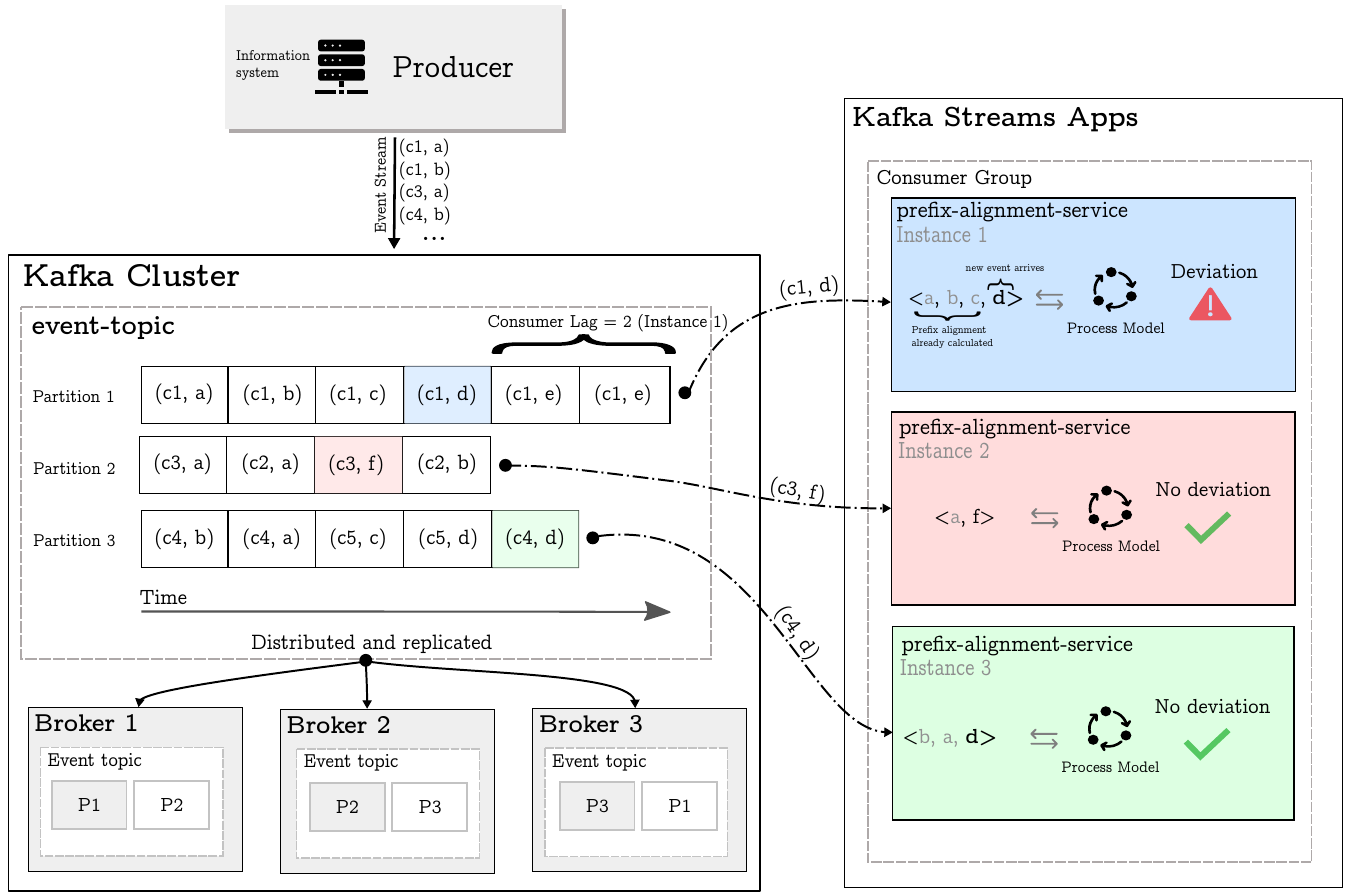}
    \caption{Architecture overview of the proposed implementation}
    \label{fig:overview_architecture}
\end{figure}

In this section, we present a horizontally scalable and fault-tolerant online conformance checking application based on \emph{Apache Kafka}~\cite{kreps2011kafka}. %and the idea of incremental prefix-alignment computation. 
In the following, we explain the system's architecture using \autoref{fig:overview_architecture}. 
%The proposed architecture allows us to perform alignment-based conformance checking in real time and at scale. 
%The corresponding service is referred to as \emph{prefix-alignment-service}. 
%First, we briefly explain the key concepts for storing and processing data with Apache Kafka. 
%Subsequently, we describe how we use these concepts in our implementation. %for the \emph{prefix-alignment-service}.

Apache Kafka is a distributed log processing/streaming system that is designed for handling high throughput of event data while maintaining low latency. 
Kafka implements the publish/subscribe pattern \cite{eugster2003many}.
The key property of a publisher is that the messages are not directed to a specific subscriber. 
Instead, a middleware-component stores the produced messages and categorizes them according to given criteria. 
Independent of the publishers, subscribers can process published messages. 
In our implementation, Kafka acts as the middleware component for event storage and offers APIs for publishing event data and subscribing to it. 
In Kafka's terminology, \emph{producers} correspond to publishers and \emph{consumers} to subscribers. 
A producer, in our case a single event stream, sends a message to one of potentially many \emph{Kafka Brokers}. % where a single Kafka server is called \emph{Broker}. 
Brokers, which are usually distributed over several physical nodes, form a \emph{Kafka cluster}. 
In \autoref{fig:overview_architecture}, we depict a Kafka cluster with three brokers. 
%Of particular relevance for the implementation is the way in which the Brokers store data and how event data can be processed. 

Kafka stores messages, i.e., the events capturing the execution of process instances, in topics. 
%The underlying logical data structure is a \emph{commit log} to which messages can only be appended. %Already produced messages are immutable objects that are persisted on a Kafka broker's disk. 
To avoid a broker holding all messages, which is impossible in some cases, topics can be split into any number of \emph{partitions}. %, with each partition containing a portion of the entire topic data. 
%Thereby, Kafka takes care of distributing partitions evenly across the available brokers. 
%Besides that it can be specified over how many instances a topic is replicated. 
In our implementation, the events from the event stream are published to the \emph{event-topic}. 
The structure of this topic is depicted in \autoref{fig:overview_architecture} within the dashed box in the Kafka Cluster. 
Further, we assume that the topic is divided into three partitions. 
Partitions are visualized as arrays where each cell contains a message, i.e., an event containing a case-id and an activity label. 
The first tuple entry refers to a case identifier of a process instance which is also the key of the message. 
The second entry is the payload consisting of the activity label. 
Kafka's default partitioning strategy ensures that messages with the same key are written to the same partition. 
Hence, events belonging to the same case are routed to the same partition.
Moreover, it is shown that the topic called \emph{event-topic} is replicated once.
Thus, there are a total of six partitions, which are evenly distributed among the brokers (indicated by boxes labeled with P1-P3).

%\subsection{Processing event data at scale}

\iffalse
We distinguish two types of scalability: vertical and horizontal scalability. 
Vertical scalability means adding hardware to an existing machine, while horizontal scalability refers to adding instances of the same application. 
The disadvantages of vertical scalability are, for example: downtime of the application during scaling, limited possibility of adding resources, no fault tolerance as the failure of one machine leads to the failure of the whole application. 
Therefore, horizontal scalability is usually the more preferable approach.
\fi
Kafka achieves horizontal scalability through consumers, who read messages according to their order in the corresponding partition. %, i.e., a Consumer always starts at offset zero. 
%For each consumer, Kafka maintains the offset of the last message consumed. 
%Consumers can be assigned to \emph{consumer groups}, which can subscribe to a set of topics. 
%As mentioned before, Kafka topics are typically partitioned. 
When a consumer group subscribes to a topic, Kafka assigns each partition to a member in the group s.t. each partition is processed by exactly one consumer. 
Thus, the various consumers can process a topic in parallel. 
Hence, the maximum degree of parallelism of an application is determined by the number of partitions of a topic.
Consequently, the way to scale a system that uses Kafka is to simply add more consumers to a group and increase the number of partitions.

% \subsection{Implementation with Kafka Streams}
\emph{Kafka Streams} is a client library used to build services that stream data from Kafka topics. 
Consider the three instances of the \emph{prefix-alignment-service} in \autoref{fig:overview_architecture}. %using the Java version of the Kafka Streams API. 
Each instance calculates prefix-alignments for the process instances whose events are written to the assigned partitions. 
Since the \emph{event-topic} is divided into three partitions, each partition is assigned to one instance. 
%Thus, prefix alignments can be calculated in parallel for three cases.
 
 \iffalse
\begin{figure}[tb]
  \centering
  \includegraphics[width=1\textwidth]{topology.pdf}
  \caption{Stream transformation steps}
  \label{fig:topology}
\end{figure}
 
Figure \ref{fig:topology} shows the chain of operators applied to the event stream by the \emph{prefix-alignment-service}. 
First of all the event stream published to Kafka from some information system is grouped by the case identifier of the respective process instance. 
The grouped stream is then aggregated so that for each case of a process an entry is stored in a local state store. This store contains the synchronous product net and the status of the search \cite{schuster2020online}. 
At each incoming event of a certain process instance, a new prefix alignment is calculated and the corresponding aggregate is updated. 
The resulting event stream of aggregates can now be further transformed. Exemplary this is shown with the operator \emph{mapValues} in figure \ref{fig:topology}, which could e.g. yield an output stream that contains the current optimal prefix alignment for each case identifier.
\fi

\subsection{Direct Synchronizing}
\label{sec:direct_sync}
In this section, we introduce an extension of the proposed implementation to improve the calculation time.
In particular, we explain how we can skip shortest path searches in cases where we can immediately return the shortest path on an extended search space upon receiving a new event.

The idea of direct synchronizing is to skip shortest path searches in cases where we can simply extend the previous calculated prefix-alignment by a synchronous move on the newly observed activity.
Hence, we execute the first and second step as described in~\autoref{sec:incr_search_basics}. 
Next, we check if we can add a synchronous move without executing a shortest path search.
We depict this check in \autoref{alg:direct_sync}.

\begin{algorithm}[h]
    \scriptsize
	\footnotesize
	\caption{Direct synchronizing}
	\label{alg:direct_sync}
	\SetKwInOut{Input}{input}
	%\SetKwInOut{Output}{output}
	\Input{$\petriNet{=}(\places,\transitions,\pnArcs,\labelFunc)$ \tcp*[h]{reference process model, i.e., a sound WF-net}\\
	$ \sigma{=}\sigma' {\cdot}\langle a \rangle{\in}\mathcal{A}^* $ with 
	$a{\in}\mathcal{A}$ \tcp*[h]{extended trace}\\
	
	$\prefixAlignment{\in}\big((\univAct{\cup}\{\gg\})\times(\transitions{\cup}\{\gg\})\big)^*$ \tcp*[h]{previous prefix-alignment for $\sigma'$ and $N$} 
	}
	\Begin{
	    \nl $S{=}(P^S,T^S,F^S,\lambda^S)$ with $M_i \gets $ create/extend SPN for $\petriNet$ and $\sigma$\; \label{alg:line:spn_extension}
	    
	    \nl $\sigma_{\prefixAlignment} \gets $ extract sequence of transitions from $T^S$ corresponding to $\prefixAlignment$\; \label{alg:line:sequence}
        
        \nl let $M'$ s.t. $(S,M_i)\xrightarrow{\sigma_{\prefixAlignment}}(S,M')$\;
        \label{alg:line:last_marking}
        
        \nl \For(\tcp*[h]{iterate over transitions from SPN $S$}){$(t',t) {\in} T^S$}{
            \nl \If{$ (S,M')[(t',t)\rangle \land \lambda^S\big((t',t)\big){=} (a,a)$ }{\label{alg:line:sync_transition_enabled}
                \tcp{transition $(t',t)$ is enabled and represents a synchronous move on the new activity $a$}
                \nl\Return $\prefixAlignment \cdot \Big\langle \big(a,(t',t)\big) \Big\rangle $\tcp*[r]{append sync. move to prefix-alignment}\label{alg:line:add_sync_move}
            }\label{alg:line:optimal}
        } 
        \nl apply standard approach\tcp*[r]{direct synchronizing is not possible}
		
	}			
\end{algorithm}

As input we assume a reference process model $N$, the extended trace $\sigma$ where $\sigma'$ represents the previous trace for the corresponding case, and the previously calculated prefix-alignment $\prefixAlignment$ of the trace $\sigma'$ and $N$.
First, we create/extend the SPN for the extended trace and the reference process model (\autoref{alg:line:spn_extension}).
Next, we translate the previous prefix-alignment $\prefixAlignment$ to a sequence of transitions in the SPN $S$ (\autoref{alg:line:sequence}).
Note that this is always possible because every \mbox{(prefix-)}alignment corresponds to a sequence of transitions in the corresponding SPN.
Moreover, since we always extend the SPN upon receiving a new event, such sequence in the extended SPN exists that corresponds to $\prefixAlignment$. 
Given the sequence of transitions $\sigma_{\prefixAlignment}$, we determine the state $M'$ where the previous search stopped (\autoref{alg:line:last_marking}).
Next, we check if we can directly execute a transition representing a synchronous move for the new activity $a$ in this state (\autoref{alg:line:sync_transition_enabled}).
If true, we simply extend the previous prefix-alignment by a synchronous move and return (\autoref{alg:line:add_sync_move}).
Otherwise, we apply the standard approach, i.e., the third and fourth step described in \autoref{sec:incr_search_basics}.

The main reason why it is beneficial to do the presented pre-check before actually continuing the shortest path search is to avoid heuristic recalculations.
Such heuristic function is part of the used heuristic search algorithm to increase search efficiency and estimates for each state the costs to reach a goal state. 
Since the search space gets extended, and also the goal states are different in each incremental search, such heuristic recalculations are needed for each incremental search.
As shown in~\cite{schuster_process_monitoring}, heuristic recalculations involve a high calculation effort.
Thus, avoiding trivial shortest path searches can potentially speed up the prefix-alignment calculation.

\subsection{Prefix Caching}
\label{sec:prefix-caching}

\begin{table}[tb]
    \scriptsize
    \caption{Conceptual idea of prefix-caching}
    \label{tab:prefix-caching}
    \centering
    \begin{tabular}{|l|c|}
    \hline
         \textbf{processed events} & \textbf{cached prefixes} \\\hline\hline
         $(c_1,a)$ & $\langle a \rangle$\\
         $\textcolor{gray}{(c_1,a),}(c_2,a)$ & $\langle a \rangle$\\
         $\textcolor{gray}{(c_1,a),(c_2,a),}(c_2,b)$ & $\langle a \rangle,\langle a,b \rangle$\\
         $\textcolor{gray}{(c_1,a),(c_2,a),(c_2,b),}(c_1,b)$ & $\langle a \rangle,\langle a,b \rangle$\\
         %$\textcolor{gray}{(c_1,a),(c_2,a),(c_2,b),(c_1,b),}(c_3,a)$ & $\langle a \rangle,\langle a,b \rangle$\\
         \hline
    \end{tabular}
\end{table}

In this section, we introduce prefix-caching for the incremental prefix-alignment approach.
In an online environment, where multiple process instances of the same process are running, it is likely that the sequences of activities performed is similar to some degree.
Thus, one wants to avoid recalculating prefix-alignments for event sequences that were already observed in the past.
By applying prefix-caching, we avoid solving identical shortest path problem multiple times for process instances that share a certain prefix.

For instance, assume the event stream $\big\langle (c_1,a),(c_2,a),(c_2,b),\allowbreak(c_1,b),\allowbreak\dots \big\rangle$.
Consider \autoref{tab:prefix-caching}, where we show the cached prefixes while processing the sample event stream.
Per cached prefix we save intermediate search results representing the current state of the search, i.e., already explored states from the search space. 
%Hence, the open- and closed set represents the current state of the search. 
Moreover, we save the SPN and the prefix-alignment per cached prefix.
Hence, we are either able to immediately return the prefix-alignment if we have calculated it already for a given prefix or to continue the search.
In the given example (\autoref{tab:prefix-caching}), we can skip two shortest path problems and return immediately a prefix-alignment, i.e., upon receiving the event $(c_2,a)$ and $(c_1,b)$.
%Next, we introduce the architecture of the proposed implementation.

\begin{figure}[tb]
    \centering
    \includegraphics[trim=.1cm 0cm .3cm 0cm,clip,width=\linewidth]{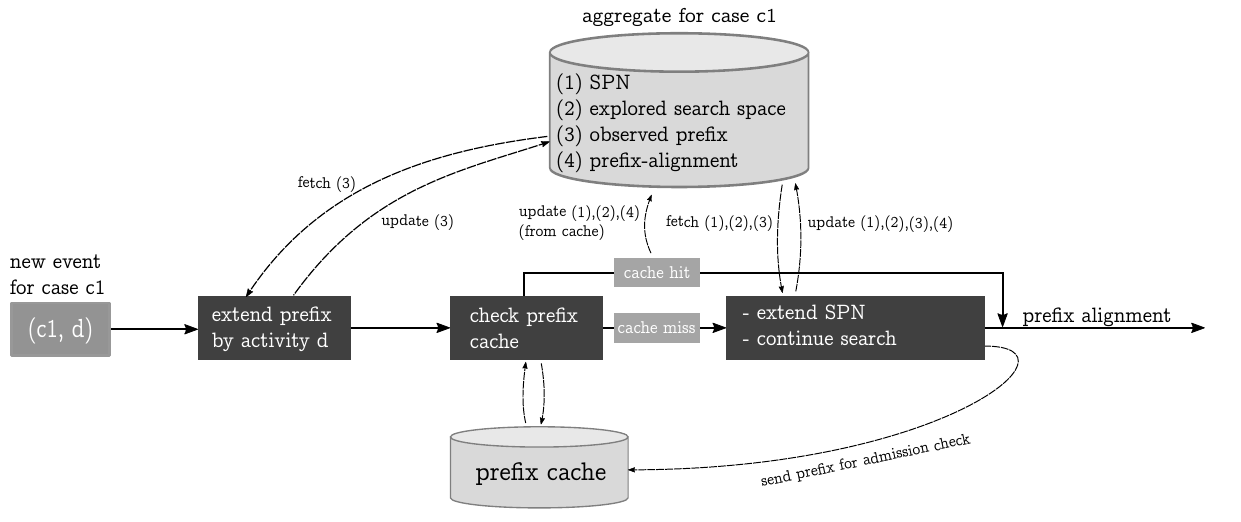}
    \caption{Overview of prefix-caching within our proposed implementation}
    \label{fig:prefix-cache}
\end{figure}

In \autoref{fig:prefix-cache}, we depict the prefix caching approach integrated within the proposed implementation.
Assume a new event $(c_1,d)$ arrives.
First, we fetch the prefix observed so far for case $c_1$ and extend it by the new activity $d$.
Note that the aggregate contains all information stored for a given case.
Next, we check the prefix-cache if we have processed the given prefix before.
If this is the case, we have a \emph{cache hit}. 
Hence, we update the SPN, the already explored search space and the prefix-alignment in the aggregate of the given case $c_1$.
Thus, we immediately return a prefix-alignment and do not solve a shortest path problem.
If we have a \emph{cache miss}, we apply the standard approach, i.e., we fetch the SPN, extend it by the new activity and continue the shortest path search from the already explored search space (\autoref{sec:incr_search_basics}).
At this stage, we can also optionally use the direct synchronizing approach presented in \autoref{sec:direct_sync} by first checking if we can directly append a synchronous move to the previous prefix-alignment.
Finally, we update the aggregate of the case and return a prefix-alignment.
Moreover, we send the new observed prefix to the prefix cache.

%TODO !!!!!!!!!!!!!!!!!!!!!!!!!!!!!!!!!!!!!!!!!!!!!!!!!!!!!!!!!!!!!!!!!!!!!!!!!!!!!!!!!!!!!!!!!!!!!!!!!!!!!!!!!!!!!!!!!!!!!!!!!!!!!!!!!!!!!!!!!!!!!!!!!!!!!!!!!!!!!!!!!!!!!!!!!!!!!!!!!!!!!!!!!!!!!!!!!!!!!!!!!!!!!!!!!!!!!!!!!!!!!!!!!!!!!!!!!!!!!!!!!!!!!!!!!!!!!!!!!!!!!!!!!!!

%The performance of a caching mechanism is determined by the \emph{hit/miss rate} and the \emph{access time}. 
%The access time is the time span between requesting the data from the cache and the arrival of the data. 
%The hit rate describes the fraction of cache requests where a prefix alignment (including it’s search state and SPN) is present in the cache. 
%Accordingly, the miss rate is the fraction of requests for which no value is stored in the cache.
Since available memory is finite, the size of the cache has to be limited.
Consequently, a cached prefix has to be replaced when a new prefix is written to the cache that has already reached its maximum capacity.
For instance, popular cache replacement algorithms are: \emph{least recently used (LRU)}, \emph{most recently used (MRU)} and \emph{least frequently used (LFU)}~\cite{podlipnig2003survey}.
Many extensions to these general strategies have been proposed that address problems such as large memory consumption for metadata. 
One example is the strategy \emph{TinyLFU}~\cite{10.1145/3149371}.
In summary, the strategy evaluates for a new cache candidate whether it should be added to the cache at the expense of deleting an already cached candidate.
For the prefix caching we use an \emph{in-process} TinyLFU cache for each node in the cluster. 
Thus, for each node in the Kafka cluster, an independent cache is maintained.

Alternatively, a \emph{distributed cache} can be used. 
However, this requires that the prefix-alignments, including the whole search status, have to be serialized and transported over the network. 
This introduces a significant overhead due to network latency and high computational cost for serialization.

\section{Experimental Evaluation}
\label{sec:evaluation}
In this section, we present an experimental evaluation of the proposed implementation and the two presented extensions.
First, we describe the experimental setup.
Subsequently, we discuss the results.

\subsection{Experimental Setup}

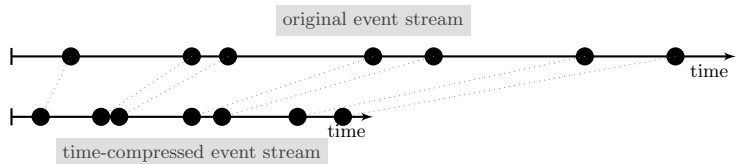
\begin{figure}[tb]
    \centering
    \resizebox{.8\textwidth}{!}{% <------ Don't forget this %
    \begin{tikzpicture}
        
        \node[darkgray,fill=lightgray!50] at (6,.6) {original event stream};
        \draw[line width=1pt, |->, >=latex'](0,0) -- coordinate (x axis) (12,0) node[below left] {time}; 
        \fill[black] (1,0) circle (0.15);
        \fill[black] (3,0) circle (0.15);
        \fill[black] (3.6,0) circle (0.15);
        \fill[black] (6,0) circle (0.15);
        \fill[black] (7,0) circle (0.15);
        \fill[black] (9.5,0) circle (0.15);
        \fill[black] (11,0) circle (0.15);
        
        \draw[dotted,gray,->] (1,0) to (1*.5,-1);
        \draw[dotted,gray,->] (3,0) to (3*.5,-1);
        \draw[dotted,gray] (3.6,0) to (3.6*.5,-1);
        \draw[dotted,gray] (6,0) to (6*.5,-1);
        \draw[dotted,gray] (7,0) to (7*.5,-1);
        \draw[dotted,gray] (9.5,0) to (9.5*.5,-1);
        \draw[dotted,gray] (11,0) to (11*.5,-1);
        
        \node[darkgray,fill=lightgray!50] at (3,-2.1+.5) {time-compressed event stream};
        \draw[line width=1pt, |->, >=latex'](0,-1) -- coordinate (x axis) (12*.5,-1) node[below left] {time}; 
        \fill[black] (1*.5,-1) circle (0.15);
        \fill[black] (3*.5,-1) circle (0.15);
        \fill[black] (3.6*.5,-1) circle (0.15);
        \fill[black] (6*.5,-1) circle (0.15);
        \fill[black] (7*.5,-1) circle (0.15);
        \fill[black] (9.5*.5,-1) circle (0.15);
        \fill[black] (11*.5,-1) circle (0.15);
    \end{tikzpicture}
    }
    \caption{Visualization of a time-compressed event stream. Each dot represents an event. The original event stream is compressed by $50\%$ s.t. that relative time distance between the events is remained w.r.t. the original event stream}
    \label{fig:event_stream_compression}
\end{figure}

In the conducted experiments, we use publicly available, real-life event logs \cite{van_dongen_2020,bpi_ch_19,van_dongen_2017,de_leoni_mannhardt_2015}.
From the event logs, we generate an event stream by emitting the events according to their timestamps.
Since the used event logs cover a large time span, we apply time-compression as visualized in \autoref{fig:event_stream_compression}.
Moreover, we discovered a reference model for each event log with the Inductive Miner~\cite{DBLP:conf/bpm/LeemansFA13}.
We conducted experiments for the two presented extensions presented in \autoref{sec:incr_search}:
%Hence, four versions of the application are tested:
%\vspace{-.2cm}
\begin{enumerate}
    \item \textbf{PL:} plain version (\autoref{sec:incr_search_basics} and \autoref{sec:implementation})
    \item \textbf{DS:} extension direct synchronizing (\autoref{sec:direct_sync})
    \item \textbf{CA:} extension prefix caching (\autoref{sec:prefix-caching})
    \item \textbf{DSC:} both extensions, i.e., direct synchronizing and prefix caching
\end{enumerate}
%\vspace{-.2cm}
We use a five node Kafka cluster with each broker running on a separate physical machine.
Consider \autoref{tab:hardware_config} for detailed specifications.
In cases where we use prefix-caching, we set the cache size to 100 prefixes per instance.

\begin{table}[tb]
    \caption{Specifications of the hardware used in the experimental setup}
    \label{tab:hardware_config}
    \centering
    \scriptsize
    \begin{tabular}{|c|c|c|}
    \hline
        Node & Memory &  CPU  \\
        \hline\hline
          1-2  & 128 GB RAM  &  2x Xeon 5115 Gold @ 2.40 GHz base    \\
          %2  & 128 GB RAM  &  2x Xeon 5115 Gold @ 2.40 GHz base                              \\
          3-5  & 512 GB RAM  &    2x Xeon 5115 Gold @ 2.40 GHz base\\                            
          %Node 4  & 512 GB RAM  &  2 x Xeon 5115 Gold @ 2.40 GHz base                              \\
          %Node 5  & 512 GB RAM  &   2 x Xeon 5115 Gold @ 2.40 GHz base     
          \hline
    \end{tabular}
\end{table}

\subsection{Results}

\begin{figure}[tb]
    \centering
    \includegraphics[width=.65\textwidth]{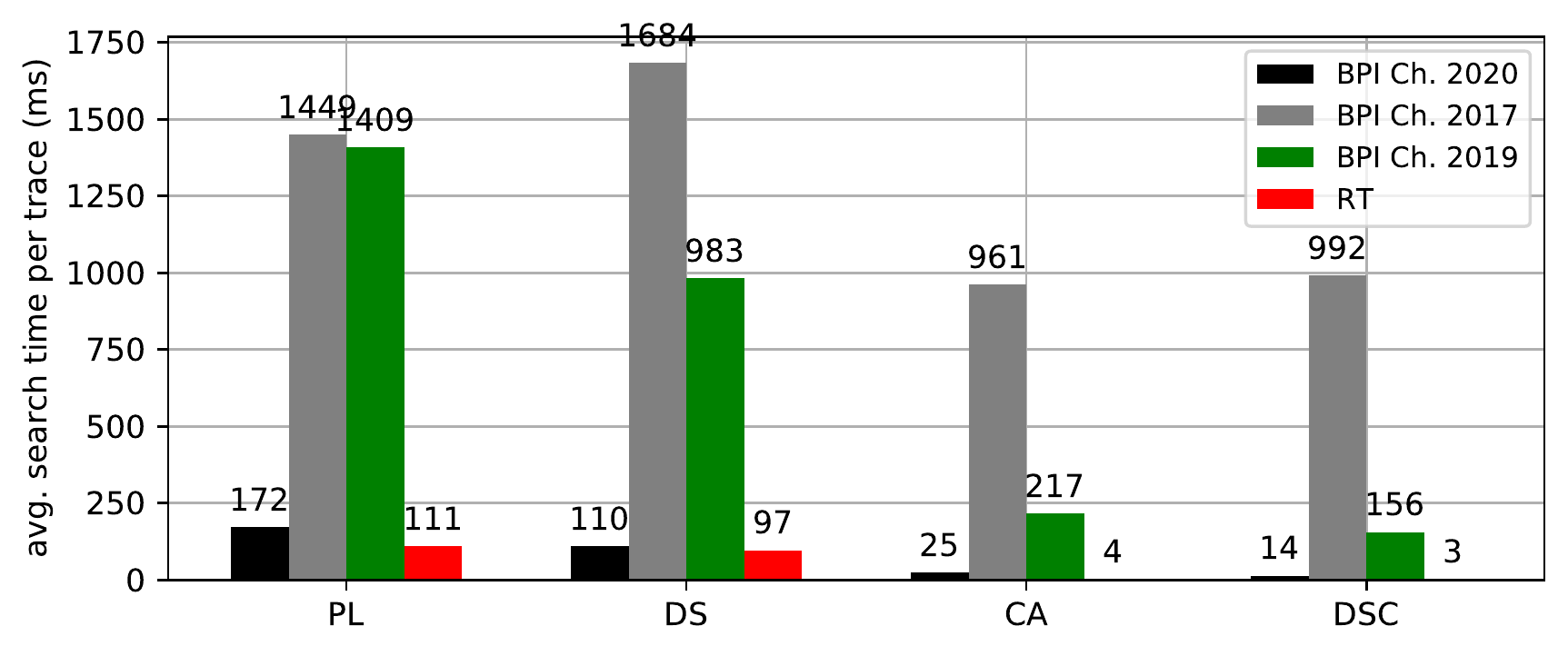}
\caption{Average computation time (ms) per trace} 
\label{fig:computation_time}
\end{figure}

\begin{figure}[tb]
  \begin{subfigure}{0.24\textwidth}
    \includegraphics[trim=.28cm 0cm .25cm .1cm,clip,width=\linewidth]{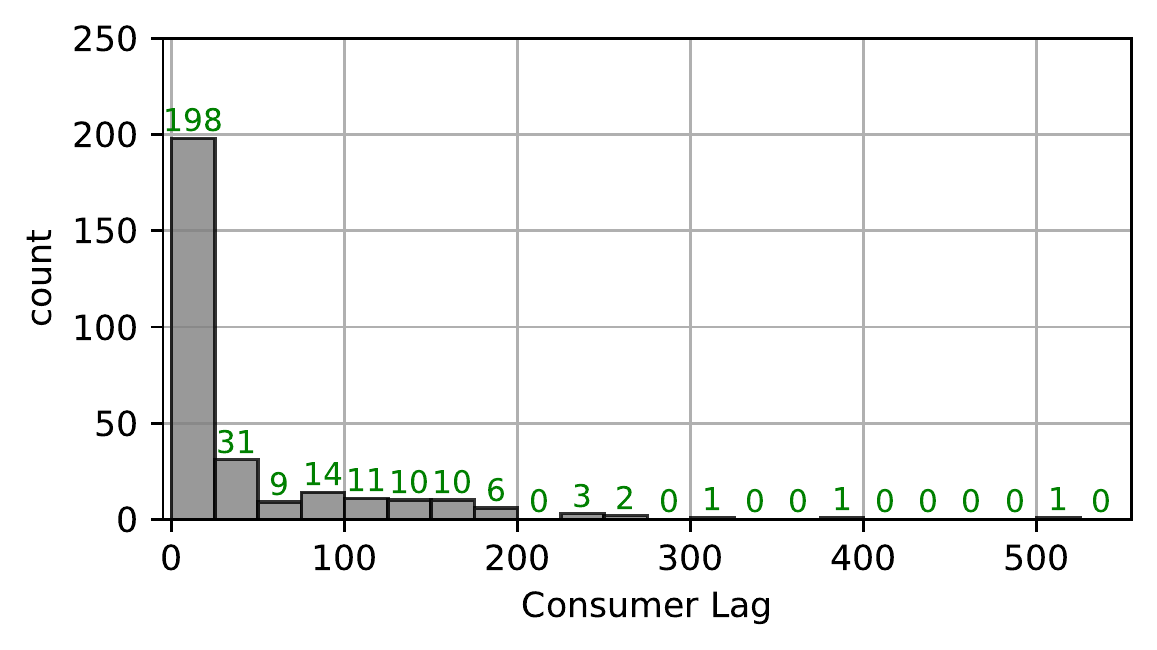}
    \caption{PL} \label{fig:1a}
  \end{subfigure}%
  \hfill
  \begin{subfigure}{0.24\textwidth}
    \includegraphics[trim=.28cm 0cm .25cm .1cm,clip,width=\linewidth]{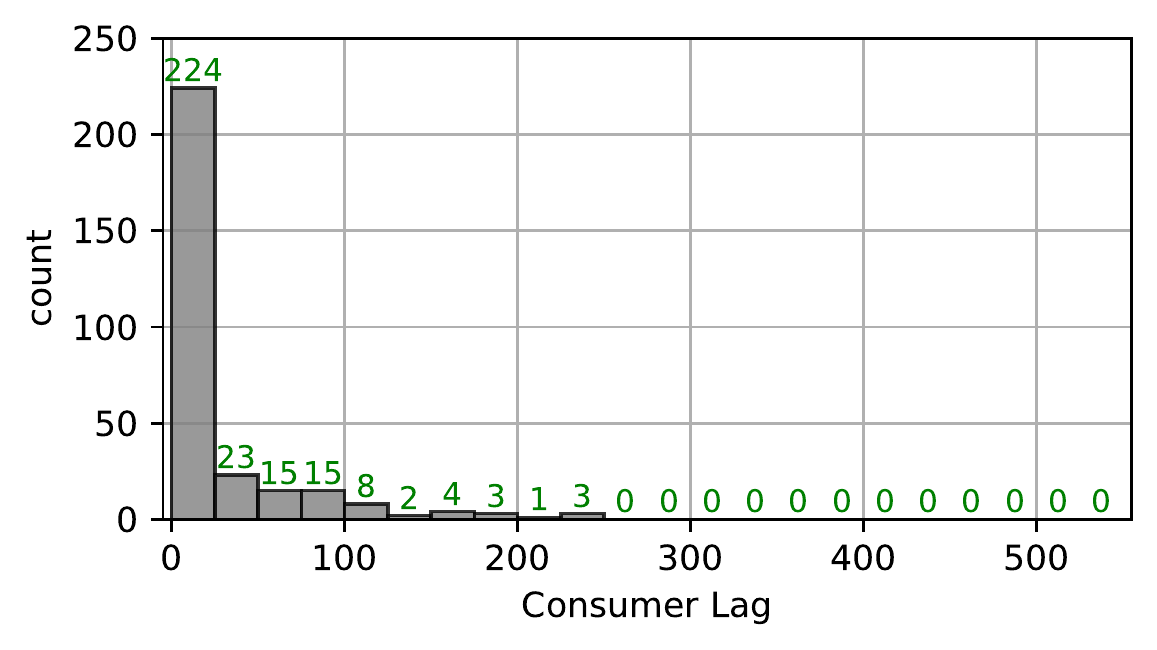}
    \caption{DS} \label{fig:1b}
  \end{subfigure}%
  \hfill
  \begin{subfigure}{0.24\textwidth}
    \includegraphics[trim=.28cm 0cm .25cm .1cm,clip,width=\linewidth]{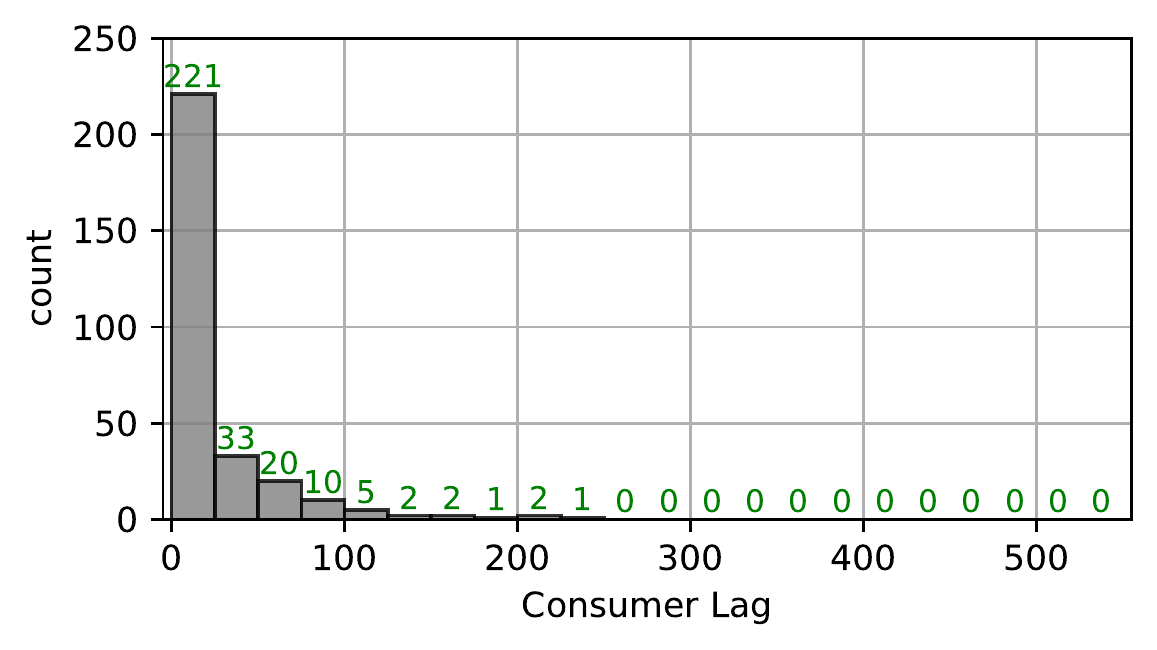}
    \caption{CA} \label{fig:1c}
  \end{subfigure}
  \hfill
  \begin{subfigure}{0.24\textwidth}
    \includegraphics[trim=.28cm 0cm .25cm .1cm,clip,width=\linewidth]{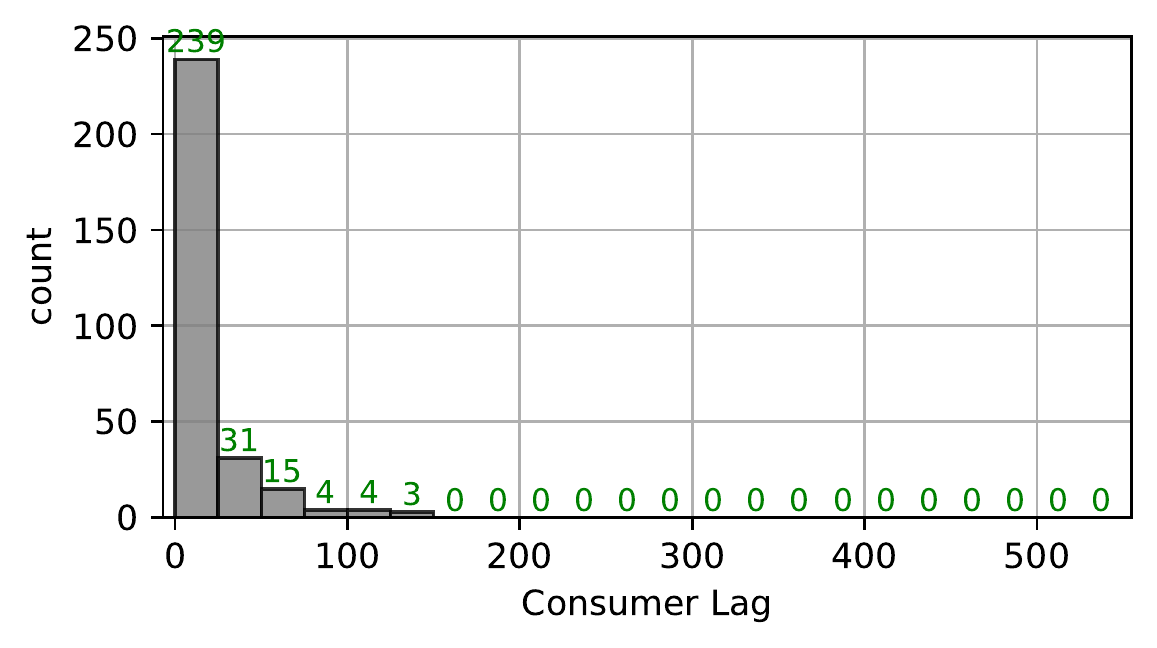}
    \caption{DSC} \label{fig:1d}
  \end{subfigure}

\caption{Distribution of consumer lag count for the four algorithm-variants using the BPI Ch. 2020 domestic log~\cite{van_dongen_2020} with a replay time of 10 minutes} 
\label{fig:consumer_lag}
\end{figure}

\iffalse
\begin{figure}[tb]
  \begin{subfigure}[t]{0.4\textwidth}
    \includegraphics[trim=.2cm .2cm .2cm .1cm,clip,width=\linewidth]{lag_histo_cds_45min.pdf}
    \caption{Consumer lag}     
  \end{subfigure}%
  \hfill
  \begin{subfigure}[t]{0.45\textwidth}
    \includegraphics[trim=.2cm .2cm .2cm .1cm,clip,width=\linewidth]{BPIChallenge2020domestic/produced_consumed_40min.pdf}
    \caption{Produced and consumed events}   
    \label{fig:produced-consumed}
  \end{subfigure}%
  \caption{Results for BPI Ch. 2020 applying DSC with replay time of 40 minutes}
  \label{fig:bpi_ch_2020_40_min}
\end{figure}
\fi

Consider \autoref{fig:computation_time} showing the average computation time per trace for the four different versions.
For all tested event logs, we observe that applying the two proposed extensions, i.e., DSC, leads to a significant speed-up of the calculation compared to the original PL version.
Moreover, we observe synergetic effects of both proposed extensions, i.e., DS and CA compared to DSC, for all logs except for BPI Ch. 2017.
Interestingly, we observe that for the BPI Ch. 2017 log, the extension DS performs worse than PL. 
This can be explained by the fact that for this event log, direct synchronization could be applied only in a few cases. 
Hence, the additional check, i.e., \autoref{alg:line:spn_extension}-\ref{alg:line:optimal} in \autoref{alg:direct_sync}, is causing the higher computation time.
However, for the other tested event logs, both extensions significantly reduce the calculation time.

In \autoref{fig:consumer_lag}, we show the distribution of the consumer lag for the four variants.
We observe that the extensions, especially DSC including both extensions, lead to a significant improvement of the consumer lag, i.e., number of queued states for consumption (reconsider \autoref{fig:consumer_lag}).  
Note that we replay the log in only 10 minutes whereas the original log spans 890 days, i.e., the time difference between the earliest and latest event in the event log.
Regarding the consumer lag distribution, we observe similar results for the other tested event logs.

\iffalse
In \autoref{fig:bpi_ch_2020_40_min}, we show the consumer lag for the BPI Ch. 2020 log and the variant DSC when using a replay time of 40min.
In comparison to the replay time of 10min (\autoref{fig:consumer_lag}), we observe a significant lower consumer lag. 
Considering \autoref{fig:produced-consumed}, we observe that our implementation is capable of processing the events emitted over time with little to no delay.
\fi

 \begin{figure}[h]
  \begin{subfigure}[t]{0.49\textwidth}
  \centering
    \includegraphics[trim=.3cm .3cm .2cm .25cm,clip,width=\linewidth]{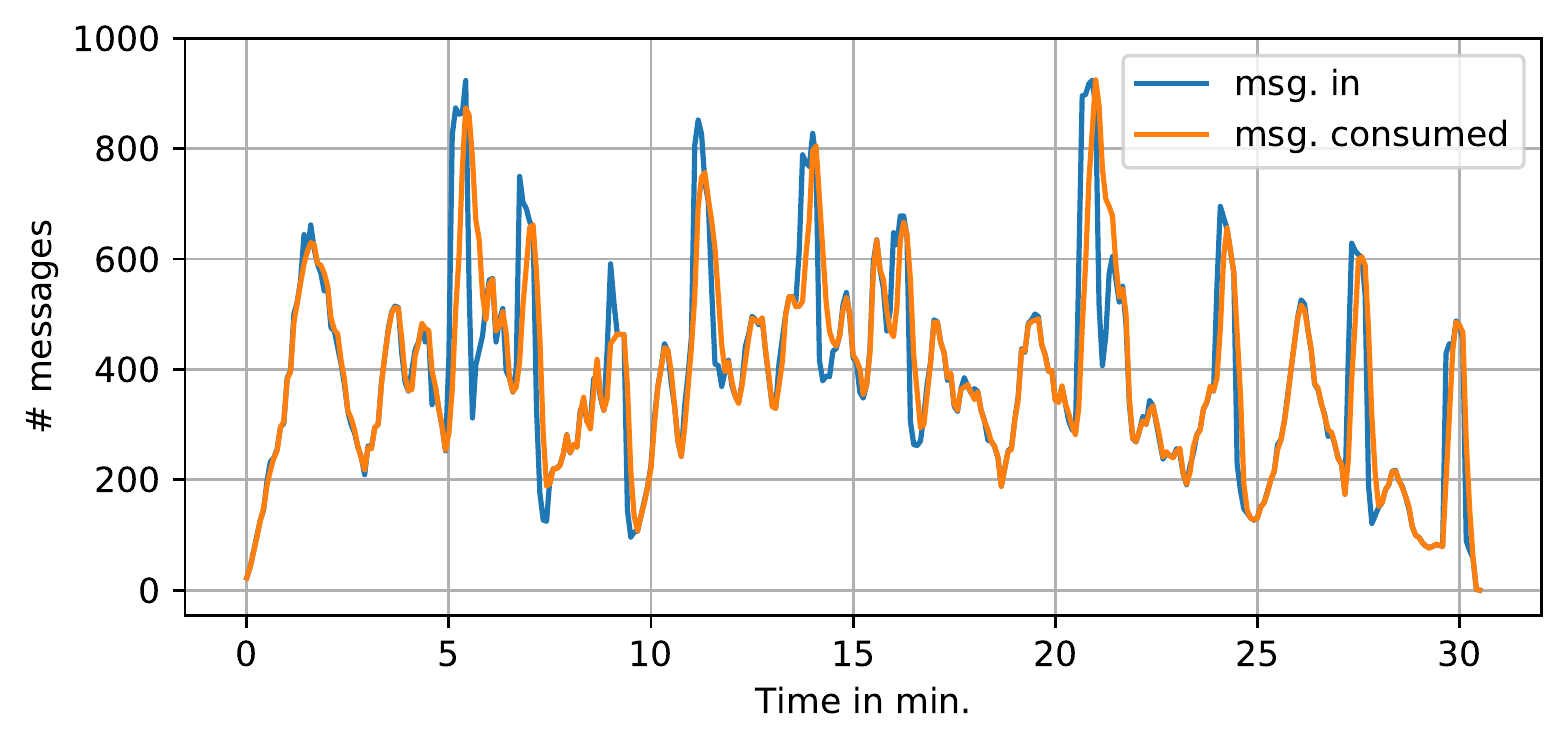}
    \caption{PL}     
  \end{subfigure}%
  \hfill
  \begin{subfigure}[t]{0.49\textwidth}
    \centering
    \includegraphics[trim=.3cm .3cm .2cm .25cm,clip,width=\linewidth]{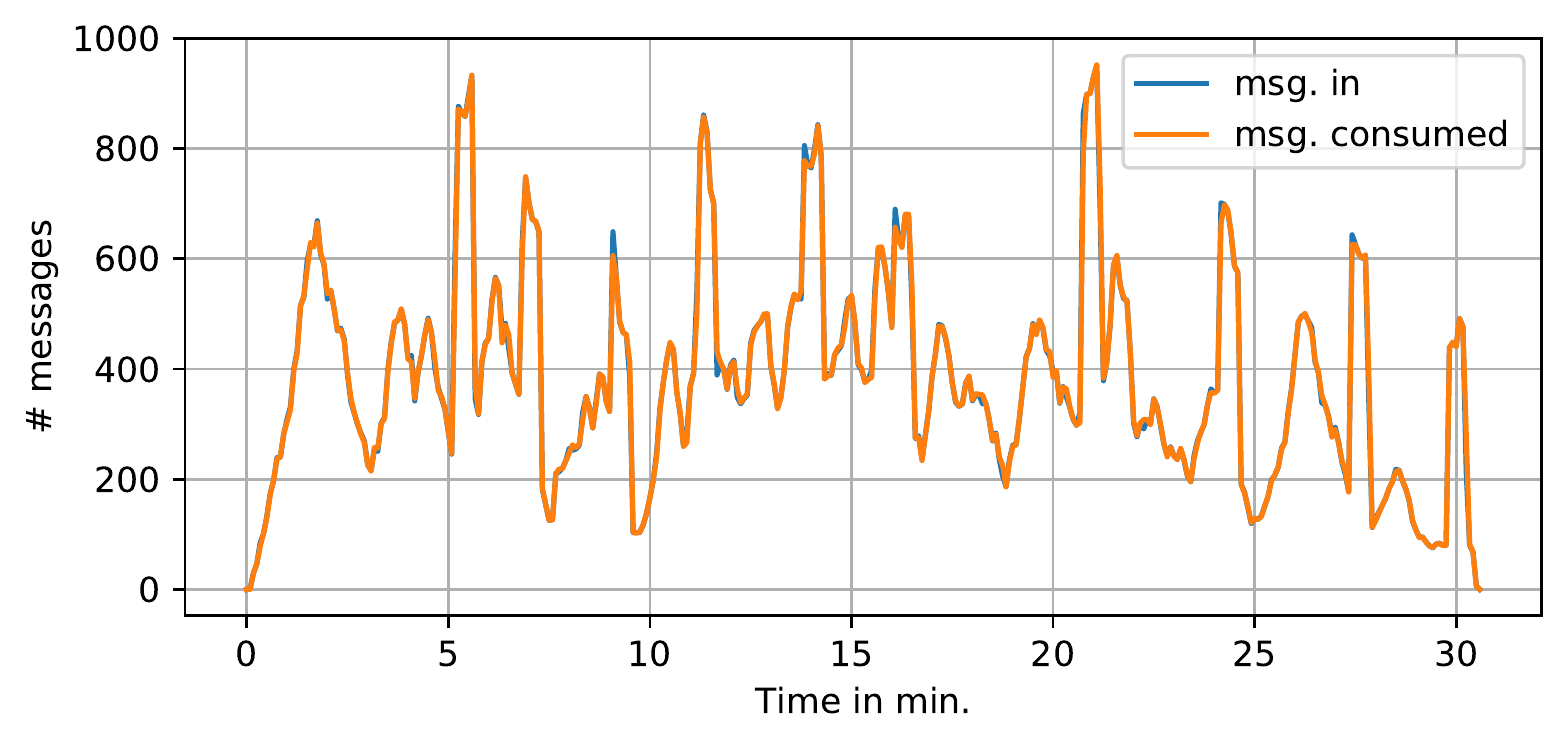}
    \caption{DSC}   
    %\label{fig:produced-consumed}
  \end{subfigure}%
  \caption{\emph{Messages in} vs. \emph{messages consumed} for RT event log}
  \label{fig:road_traffic_msg_in_out}

  \begin{subfigure}[t]{0.49\textwidth}
  \centering
    \includegraphics[trim=.3cm .3cm .2cm .25cm,clip,width=\linewidth]{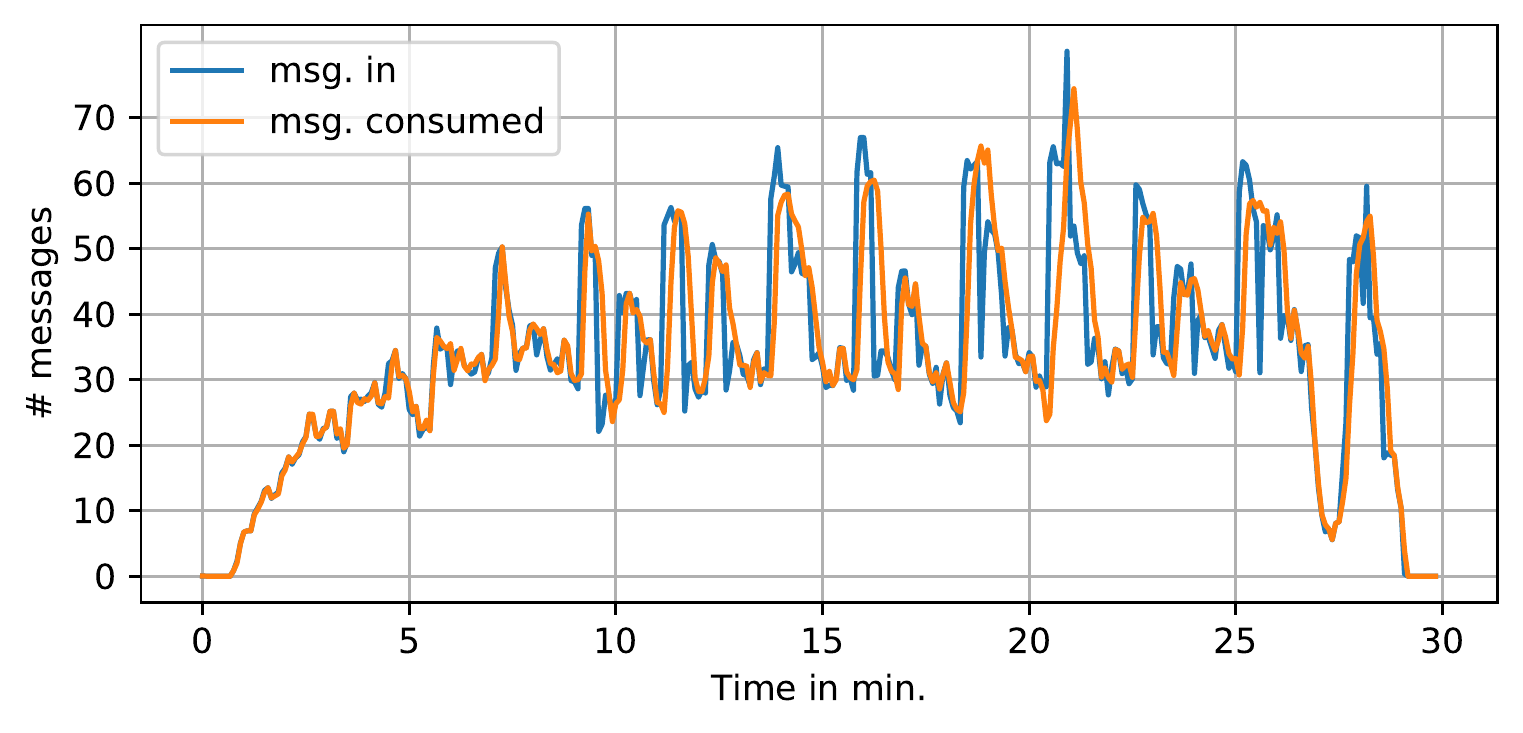}
    \caption{PL}     
  \end{subfigure}%
  \hfill
  \begin{subfigure}[t]{0.49\textwidth}
  \centering
    \includegraphics[trim=.3cm .3cm .2cm .25cm,clip,width=\linewidth]{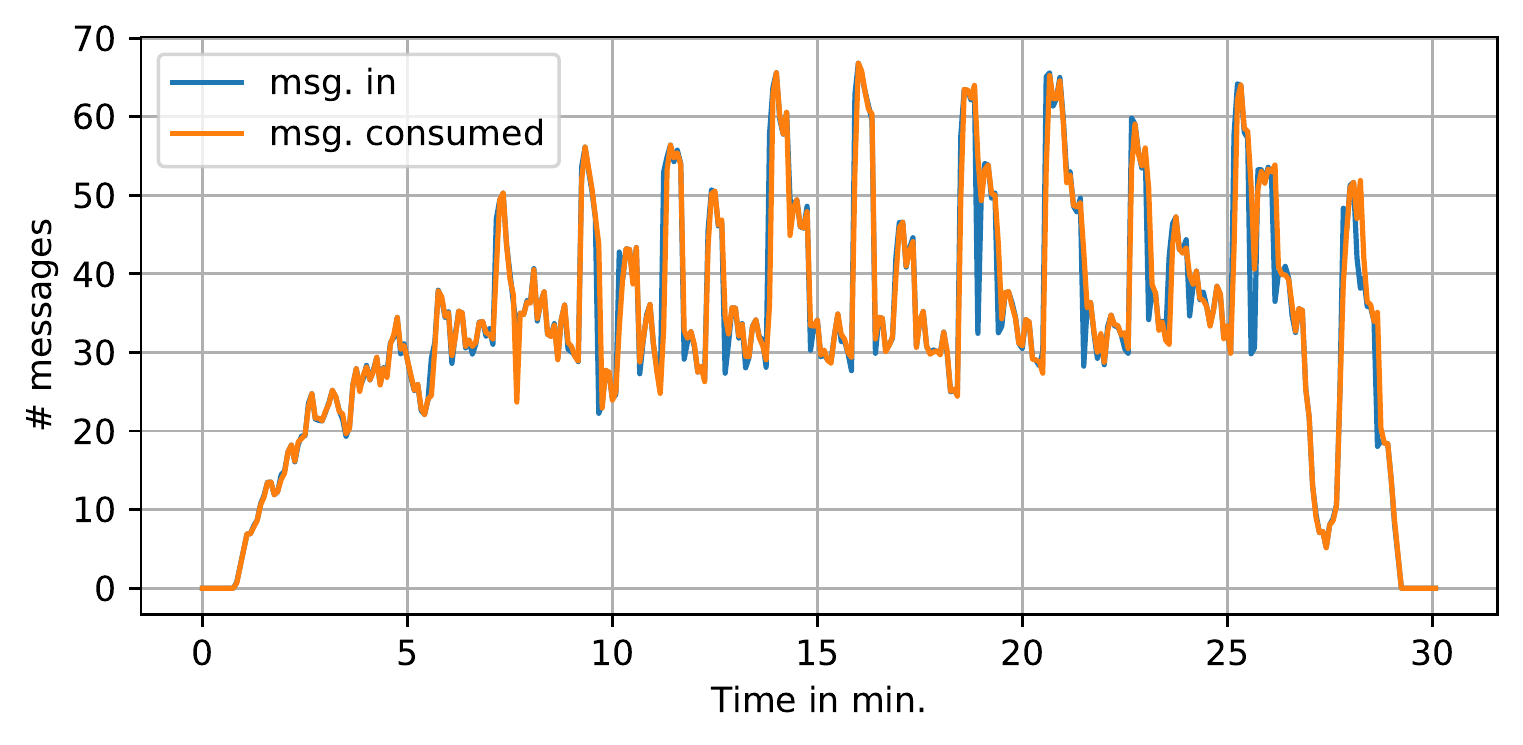}
    \caption{DSC}   
    %\label{fig:produced-consumed}
  \end{subfigure}%
  \caption{\emph{Messages in} vs. \emph{messages consumed} for BPI Ch. 2019 event log}
  \label{fig:bpi_2019_msg_in_out}

  \begin{subfigure}[t]{0.49\textwidth}
  \centering
    \includegraphics[trim=.3cm .3cm .2cm .25cm,clip,width=\linewidth]{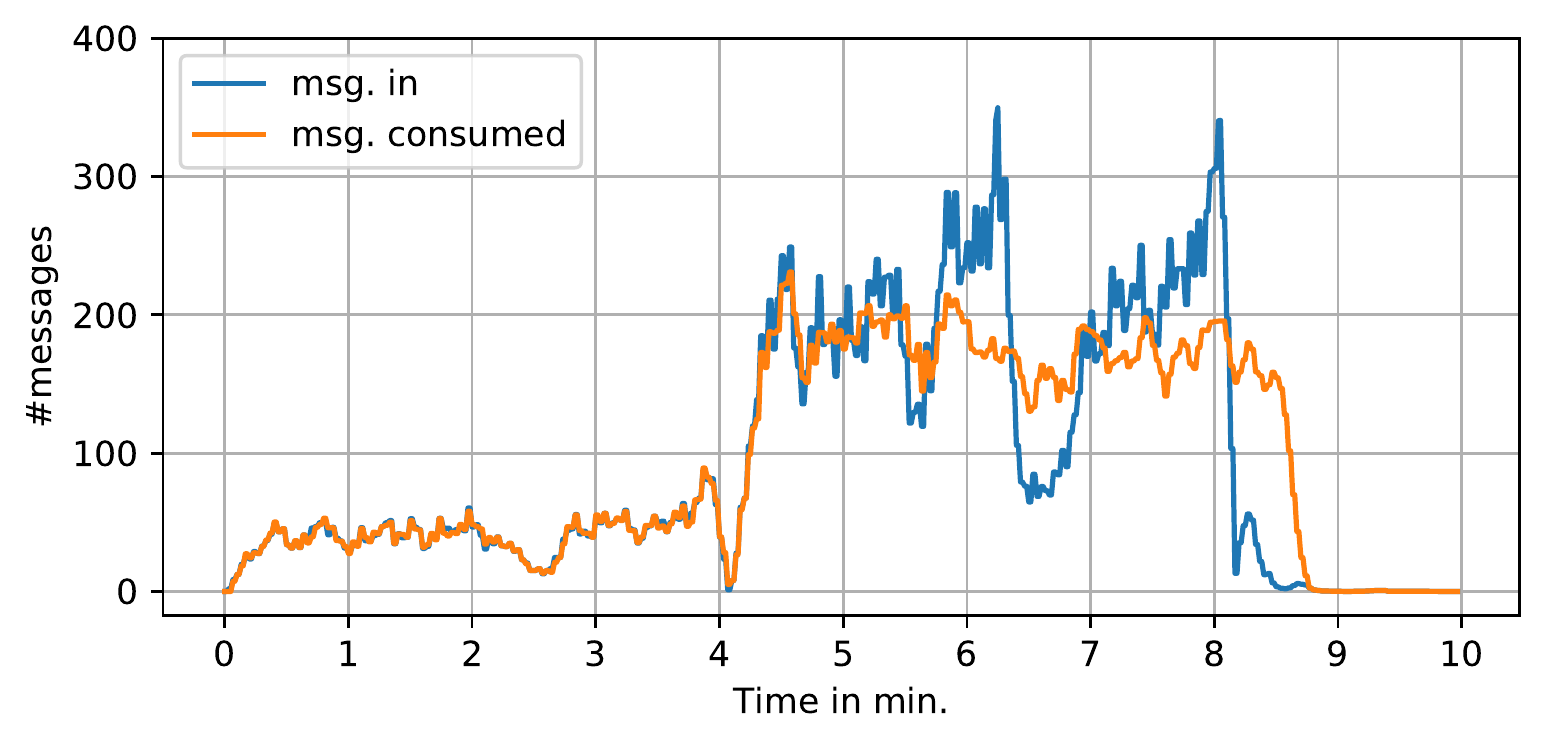}
    \caption{PL}     
  \end{subfigure}%
  \hfill
  \begin{subfigure}[t]{0.49\textwidth}
  \centering
    \includegraphics[trim=.3cm .3cm .2cm .25cm,clip,width=\linewidth]{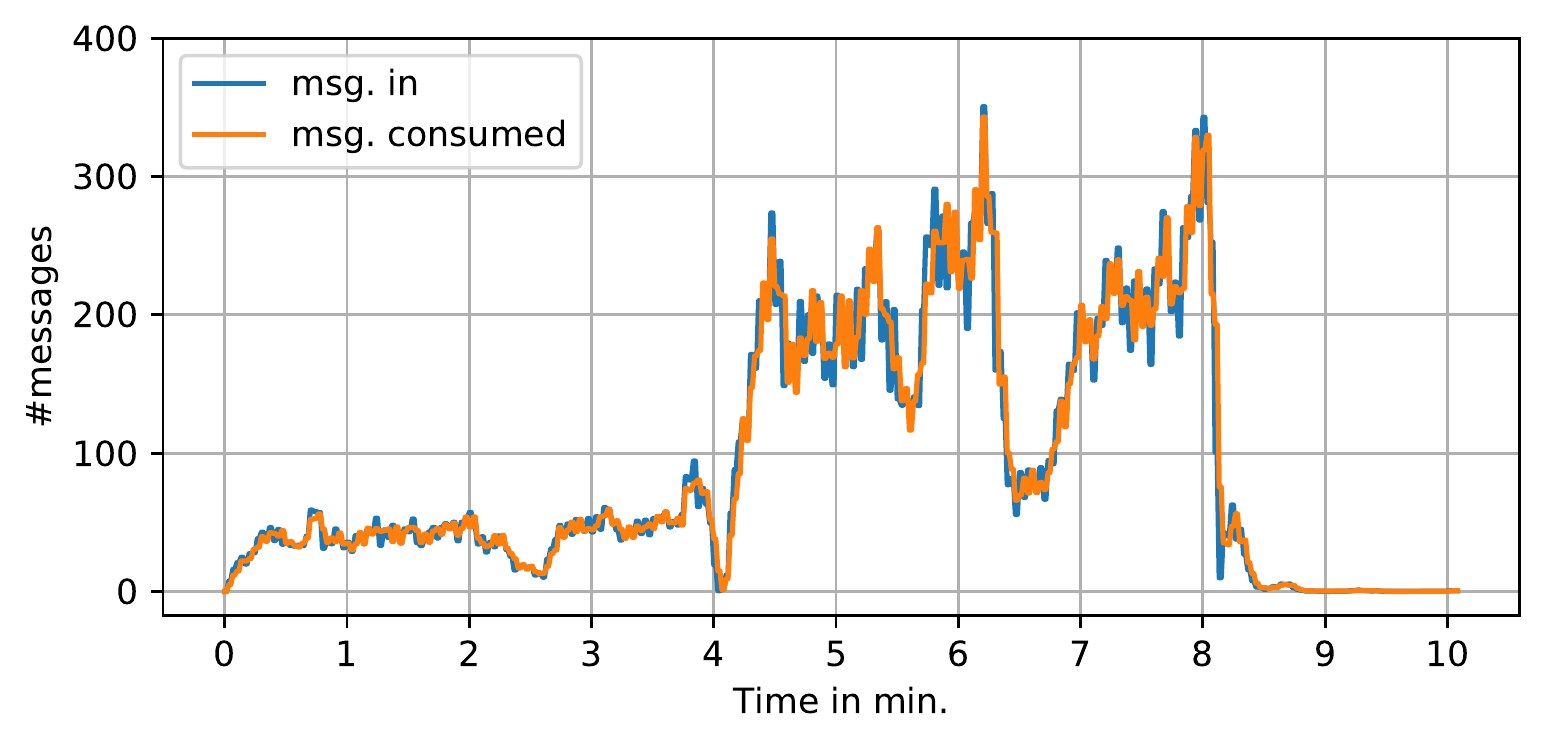}
    \caption{DSC}   
    %\label{fig:produced-consumed}
  \end{subfigure}%
  \caption{\emph{Messages in} vs. \emph{messages consumed} for BPI Ch. 2020 event log}
  \label{fig:bpi_2019_msg_in_out}

  \begin{subfigure}[t]{0.49\textwidth}
  \centering
    \includegraphics[trim=.3cm .3cm .2cm .25cm,clip,width=\linewidth]{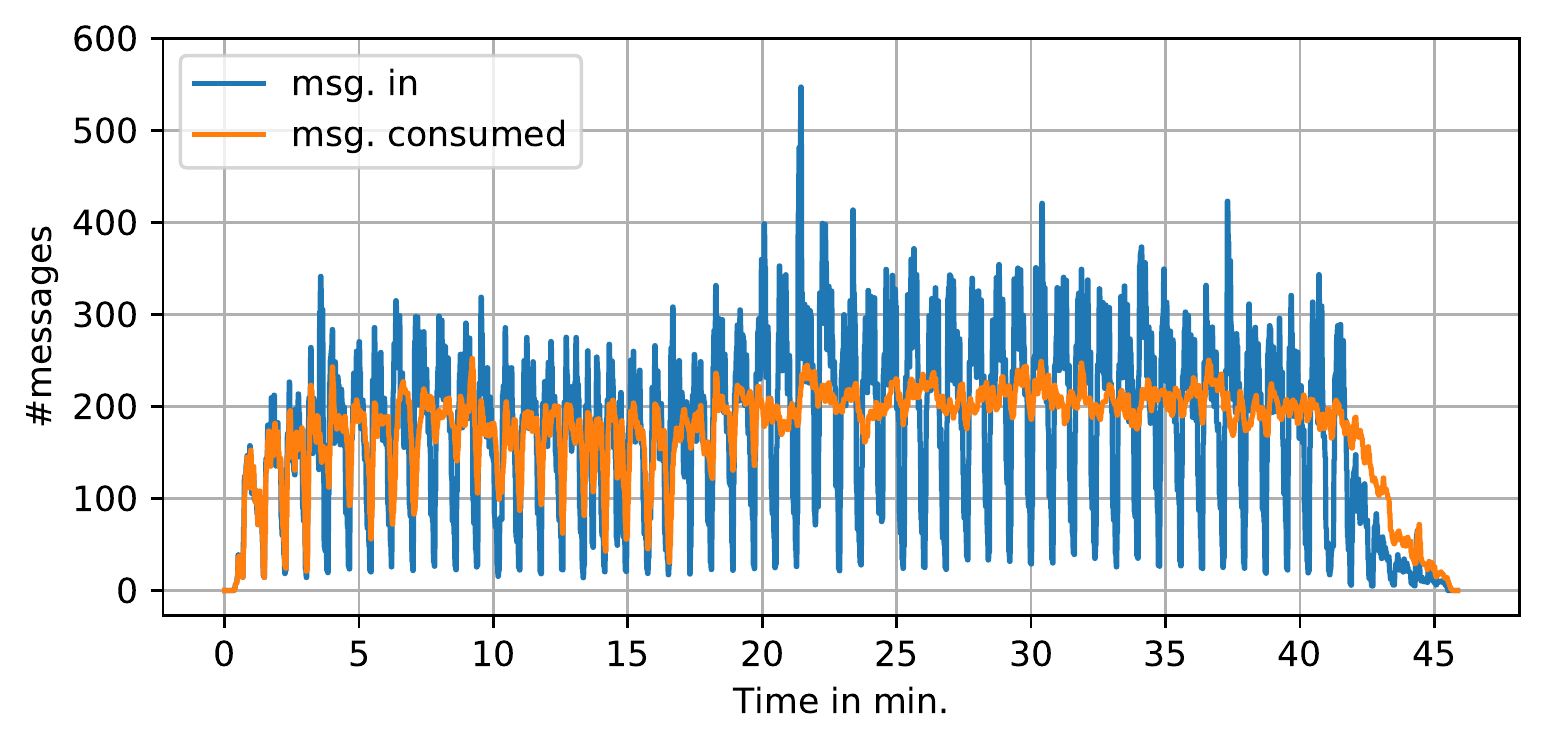}
    \caption{PL}     
  \end{subfigure}%
  \hfill
  \begin{subfigure}[t]{0.49\textwidth}
  \centering
    \includegraphics[trim=.3cm .3cm .2cm .25cm,clip,width=\linewidth]{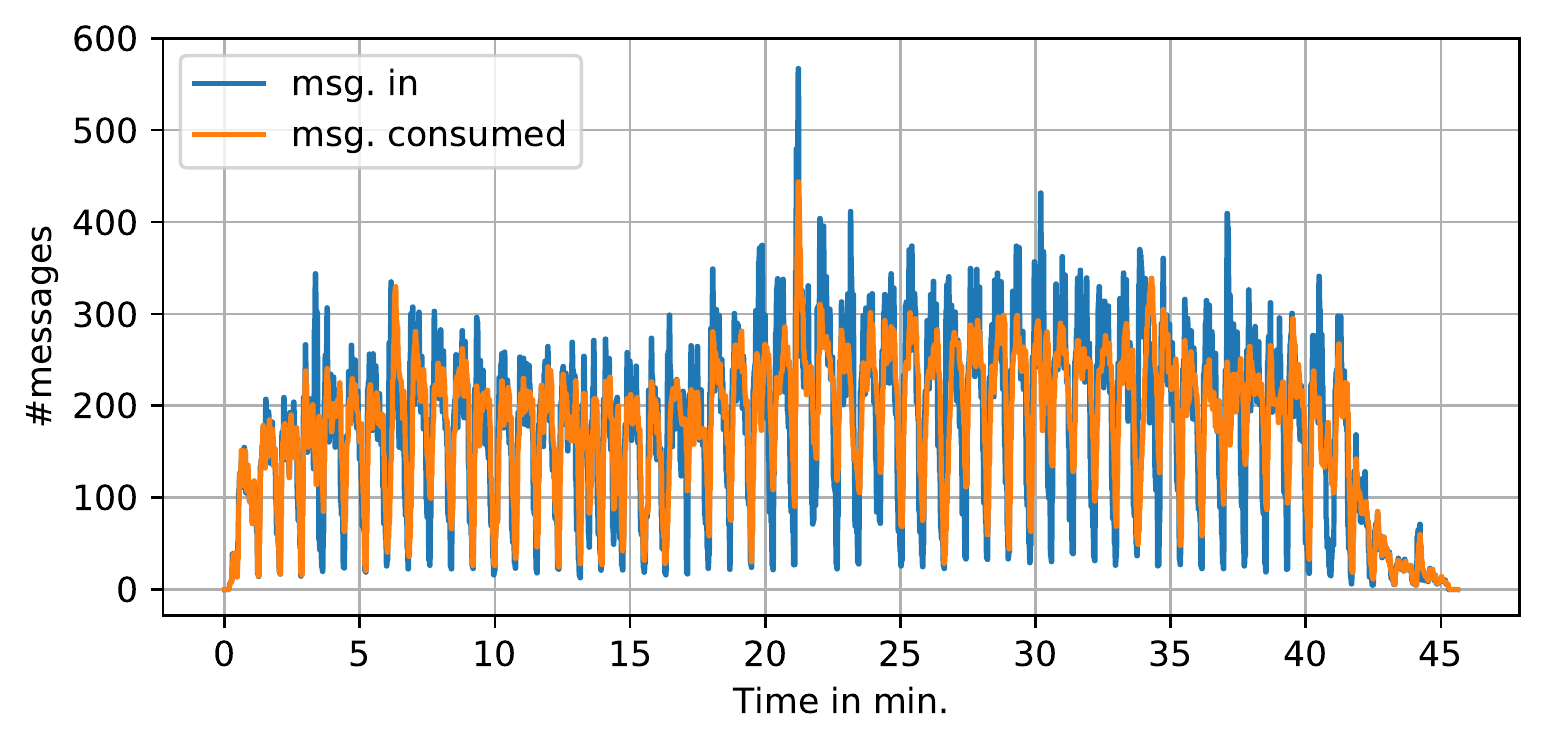}
    \caption{DSC}   
    %\label{fig:produced-consumed}
  \end{subfigure}%
  \caption{\emph{Messages in} vs. \emph{messages consumed} for BPI Ch. 2017 event log}
  \label{fig:bpi_2017_msg_in_out}
\end{figure}

%TODO
In Figures \ref{fig:road_traffic_msg_in_out}-\ref{fig:bpi_2017_msg_in_out} we depict the flow of \emph{messages in} and \emph{messages consumed} for the different event logs.
Per event log, we compare the PL version, which serves as a baseline, and the DSC version, which includes both proposed extensions.
In general, we observe that the flow of \emph{messages in}, i.e., incoming events from the stream, and the flow of \emph{messages consumed}, i.e., processed events, overlaps much more for the DSC version compared to the PL version.
Hence, the consumer lag of the DSC version is significantly lower compared to the PL version.
This shows that the two proposed extensions significantly reduce the computational effort.

\iffalse
Regarding limited memory resources, we have so far only focused on limiting the cache size.
However, since we store an aggregate for each case, which contains the SPN and the intermediate results of the search, we must consider that we cannot store infinitely many aggregates.
Although, we have never exceeded the available memory resources within the conducted experiments, which are based on real-life event logs covering a large time-span, we additionally conducted experiments where we artificially limit the available memory per instance. 
If an instance runs out of memory, we drop the oldest aggregate. 
However, we always maintain the sequence of performed activities for each case.
This allows us to start the search from scratch if a new activity is performed for a case where no intermediate search results is available anymore.
Note that this procedure still requires in theory infinite memory. 
Nevertheless, the memory needed for just storing the sequence of executed activities is low compared to the entire aggregate containing intermediate search results, i.e., explored states.
\fi
\section{Conclusion}
\label{sec:conclusion}
In this paper, we presented an implementation for scalable online conformance checking based on incremental prefix-alignment calculation.
The proposed implementation is based on Apache Kafka, a streaming platform that is widely used within industry.
Therefore, the paper offers an important basis for the industrial application of online conformance checking techniques.
%We presented an architecture for scalable online conformance checking based on incremental prefix-alignment calculation.
Moreover, we presented two extensions that significantly reduce the computational effort.
Our conducted experiments show that the proposed implementation combined with the presented extensions is capable to efficiently process real-life event streams.

Open questions that need to be addressed in future work include the problem of deciding when a process instance/case is considered to be complete.
This decision is necessary because we will eventually be required to delete aggregates, i.e. the current search state stored for each process instances (\autoref{fig:prefix-cache}), due to limited memory resources of the cluster.
However, this is a general problem within the area of online process mining and, hence, outside the scope of this paper.

\bibliographystyle{splncs04}
\bibliography{main}

\newcommand{\SortNoop}[1]{}
\begin{thebibliography}{10}
\providecommand{\url}[1]{\texttt{#1}}
\providecommand{\urlprefix}{URL }
\providecommand{\doi}[1]{https://doi.org/#1}

\bibitem{DBLP:journals/jcsc/Aalst98}
{\SortNoop{Aalst}}van~der Aalst, W.M.P.: {The Application of Petri Nets to
  Workflow Management}. Journal of Circuits, Systems, and Computers
  \textbf{8}(1),  21--66 (1998)

\bibitem{DBLP:journals/widm/AalstAD12}
{\SortNoop{Aalst}}van~der Aalst, W.M.P., Adriansyah, A.,
  {\SortNoop{Dongen}}van~Dongen, B.F.: {Replaying History on Process Models for
  Conformance Checking and Performance Analysis}. Wiley Interdisc. Rew.: Data
  Mining and Knowledge Discovery  \textbf{2}(2),  182--192 (2012)

\bibitem{DBLP:books/sp/Aalst16}
{\SortNoop{Aalst}}van~der Aalst, W.M.P.: {Process Mining - Data Science in
  Action}. Springer (2016)

\bibitem{adriansyah_2014_phd_aligning}
Adriansyah, A.: {Aligning Observed and Modeled Behavior}. Ph.D. thesis,
  Eindhoven University of Technology (2014)

\bibitem{adriansyah2013memory}
Adriansyah, A., van Dongen, B.F., van~der Aalst, W.M.: Memory-efficient
  alignment of observed and modeled behavior. BPM Center Report pp. 03--03
  (2013)

\bibitem{DBLP:conf/bpi/BurattinC17}
Burattin, A., Carmona, J.: {A Framework for Online Conformance Checking}. In:
  Proceedings BPI (2017)

\bibitem{DBLP:conf/cec/BurattinSA14}
Burattin, A., Sperduti, A., van~der Aalst, W.M.P.: {Control-flow Discovery from
  Event Streams}. In: Proceedings of the {IEEE} Congress on Evolutionary
  Computation, {CEC} 2014, Beijing, China, July 6-11, 2014. pp. 2420--2427.
  {IEEE} (2014)

\bibitem{DBLP:conf/bpm/BurattinZADC18}
Burattin, A., {\SortNoop{Zelst}}van~Zelst, S.J., Armas{-}Cervantes, A.,
  {\SortNoop{Dongen}}, van~Dongen, B.F., Carmona, J.: {Online Conformance
  Checking Using Behavioural Patterns}. In: {BPM} (2018)

\bibitem{DBLP:books/sp/CarmonaDSW18}
Carmona, J., {\SortNoop{Dongen}}van~Dongen, B.F., Solti, A., Weidlich, M.:
  Conformance Checking - Relating Processes and Models. Springer (2018)

\bibitem{bpi_ch_19}
van Dongen, B.: Dataset {BPI} challenge 2019 (2019),
  \url{https://doi.org/10.4121/uuid:d06aff4b-79f0-45e6-8ec8-e19730c248f1}

\bibitem{van_dongen_2017}
van Dongen, B.: {BPI} challenge 2017 (2017),
  \url{https://doi.org/10.4121/uuid:5f3067df-f10b-45da-b98b-86ae4c7a310b}

\bibitem{van_dongen_2020}
van Dongen, B.: {BPI} challenge 2020: Domestic declarations (2020),
  \url{https://doi.org/10.4121/uuid:3f422315-ed9d-4882-891f-e180b5b4feb5}

\bibitem{10.1145/3149371}
Einziger, G., Friedman, R., Manes, B.: Tinylfu: A highly efficient cache
  admission policy. ACM Trans. Storage  \textbf{13}(4) (Nov 2017)

\bibitem{eugster2003many}
Eugster, P.T., Felber, P.A., Guerraoui, R., Kermarrec, A.M.: The many faces of
  publish/subscribe. ACM computing surveys (CSUR)  \textbf{35}(2),  114--131
  (2003)

\bibitem{kreps2011kafka}
Kreps, J., Narkhede, N., Rao, J., et~al.: Kafka: A distributed messaging system
  for log processing. In: Proceedings of the NetDB. vol.~11, pp.~1--7 (2011)

\bibitem{DBLP:conf/bpm/LeemansFA13}
Leemans, S.J.J., Fahland, D., {\SortNoop{Aalst}}van~der Aalst, W.M.P.:
  {Discovering Block-Structured Process Models from Event Logs Containing
  Infrequent Behaviour}. In: Business Process Management Workshops - {BPM}
  2013. pp. 66--78 (2013)

\bibitem{de_leoni_mannhardt_2015}
de~Leoni, M.M., Mannhardt, F.: Road traffic fine management process (2015),
  \url{https://doi.org/10.4121/uuid:270fd440-1057-4fb9-89a9-b699b47990f5}

\bibitem{murata_1989}
Murata, T.: {Petri nets: Properties, Analysis and Applications}. Proceedings of
  the {IEEE}  \textbf{77}(4),  541--580 (Apr 1989)

\bibitem{podlipnig2003survey}
Podlipnig, S., B{\"o}sz{\"o}rmenyi, L.: A survey of web cache replacement
  strategies. ACM Computing Surveys (CSUR)  \textbf{35}(4),  374--398 (2003)

\bibitem{DBLP:journals/is/RozinatA08}
Rozinat, A., {\SortNoop{Aalst}}van~der Aalst, W.M.P.: {Conformance Checking of
  Processes Based on Monitoring Real Behavior}. Inf. Syst.  \textbf{33}(1),
  64--95 (2008)

\bibitem{schuster_process_monitoring}
Schuster, D., van Zelst, S.J.: Online process monitoring using incremental
  state-space expansion: An exact algorithm. In: BPM. pp. 147--164. Springer
  (2020)

\bibitem{DBLP:journals/ijdsa/ZelstBHDA19}
{\SortNoop{Zelst}}van~Zelst, S.J., Bolt, A., Hassani, M.,
  {\SortNoop{Dongen}}van~Dongen, B.F., {\SortNoop{Aalst}}van~der Aalst, W.M.P.:
  Online conformance checking: relating event streams to process models using
  prefix-alignments. Int. J. Data Sci. Anal.  \textbf{8}(3),  269--284 (2019)

\end{thebibliography}

\end{document}